\documentclass[final,2p,times,twocolumn]{elsarticle}

\usepackage[left=2cm,right=2cm,top=3cm,bottom=2cm]{geometry}

\usepackage{amssymb}

\usepackage[T1]{fontenc}

 \usepackage{lineno}

\usepackage[version=4]{mhchem}
\usepackage{textcomp}
\usepackage{xspace}
\usepackage{xcolor}
\usepackage{booktabs}

\usepackage{hyperref}
\hypersetup{
	colorlinks=true,         
	breaklinks=true,         
	urlcolor= orange!50!black,         
	linkcolor= orange,        
    citecolor=orange,     
    filecolor=black,      
    pagebackref=true,
}

\newcommand \wn {cm$^{-1}$\xspace}

\journal{Journal of Physical Chemistry A}

\begin{document}

\begin{frontmatter}



\title{Submillimeter-wave spectroscopy of the \ce{CH3O} radical}


\author[inst1]{Jean-Thibaut Spaniol}
\author[inst1]{Olivia Chitarra}
\author[inst1]{Olivier Pirali}
\author[inst1]{Marie-Aline Martin-Drumel\corref{cor1}}
\author[inst2]{Holger S. P. Müller\corref{cor2}}

\address[inst1]{
                Universit\'{e} Paris-Saclay, CNRS, Institut des Sciences Mol\'{e}culaires d'Orsay, 
                91405 Orsay, France
               }

\address[inst2]{
                Astrophysik/I. Physikalisches Institut, Universität zu Köln, 
                50937 Köln, Germany
               }

\cortext[cor1]{marie-aline.martin@cnrs.fr}
\cortext[cor2]{hspm@ph1.uni-koeln.de}

\begin{abstract}
The methoxy radical, \ce{CH3O}, has long been studied experimentally and theoretically by spectroscopists because it displays a weak Jahn-Teller effect in its electronic ground state, combined with a strong spin-orbit interaction. 
In this work, we report an extension of the measurement of the pure rotational spectrum of the radical in its vibrational ground state in the submillimeter-wave region (350--860~GHz).
\ce{CH3O} was produced by H-abstraction from methanol using F-atoms, and its spectrum was probed in absorption using an association of source-frequency modulation and Zeeman modulation spectroscopy.
All the observed transitions together with available literature data in $\varv = 0$ were combined and fit using an effective Hamiltonian allowing to reproduce the data at their experimental accuracy.
The newly measured transitions involve significantly higher frequencies and rotational quantum numbers than those reported in the literature ($f < 860$~GHz and $N \leq 15$ instead of 272~GHz and 7, respectively) which results in significant improvements in the spectroscopic parameters determination. The present model is well constrained and allows a reliable calculation of the rotational spectrum of the radical over the entire microwave to submillimeter-wave domain.
It can be used with confidence for future searches of \ce{CH3O} in the laboratory and the interstellar medium.
\end{abstract}


\end{frontmatter}


\section{Introduction}
\label{sec:intro}

The methoxy (\ce{CH3O}) radical, with its C$_{3\mathrm{v}}$ geometry and $^2$E electronic ground state, is an ideal test-case for studying the Jahn-Teller effect \citep[e.g.,][]{marenich2005:Model, shao2013:JahnTeller}. 
Because this effect, weak in \ce{CH3O}, is combined with a strong spin-orbit splitting \citep[e.g.,][]{yarkony1974:Geometries, bersuker2001:Modern, marenich2005:Model}, the radical has long challenged spectroscopists, experimentalists, and theoreticians alike.
Theoretical works have notably delved into the Jahn-Teller effect and vibronic coupling \citep[e.g.][]{hoeper2000:Theoretical, shao2013:JahnTeller, nagesh2014:Simulation} and the construction of a model Hamiltonian \citep[e.g.,][]{marenich2005:Model, marenich2005:Role, elhilali2009:Tensorial}.
Experimentally, the radical has been studied by laser magnetic resonance \citep{radford1977:Spectroscopic, russell1980:Analysis}, laser induced fluorescence \citep[e.g.,][]{ohbayashi1977:Emissionspectra, liu2009:CH3O}, stimulated emission pumping \citep[e.g.,][]{liu2009:CH3O, geers1994:Rotation}, electron paramagnetic resonance \citep[e.g.,][]{tsegaw2016:Electron}, and photoelectron spectroscopy \citep[e.g.,][]{engelking1978:Laser, weichman2017:Lowlying, tang2020:Threshold}; at medium to high (i.e., rotationally-resolved) resolution. The ro-vibrational spectrum of \ce{CH3O} has been recorded in the region of the $\nu_4$ band (2756--3003~\wn) in a supersonic jet expansion \cite{han2002:Highresolution, han2007:Jetcooled} while its pure rotational spectrum has been recorded at room-temperature in the 82--400~GHz region \cite{endo1984:Microwavespectrum, momose1988:Submillimeterwave} and under jet-cooled conditions in the 246--303~GHz region \cite{laas2017:Millimeter}.

Besides its purely spectroscopic interest, \ce{CH3O} is also a fundamental species for atmospheric, combustion, and interstellar chemistry. 
Its kinetics and dynamics have thus been subject to extensive investigation as well \citep[e.g.,][]{tsang1986:Chemical}. 
In the atmosphere, it is thought to be a crucial intermediate in the oxidation scheme of alkanes, in particular methane 
\cite{zellner1986:Methoxy, seinfeld2016:Atmospheric, antinolo2016:Reactivity}, and is potentially involved in smog formation through 
its reaction with \ce{O2} \cite{glasson1975:Methoxyl}. 
It is also an intermediate in the combustion of  methanol and methanol-containing fuels \cite{russell1980:Analysis, dames2013:Master}. 
The radical was first detected in the interstellar medium toward the B1-b cold core \cite{cernicharo2012:CH3O} 
and has since then been also observed in four prestellar cores \cite{bacmann2016:Origin} and toward the L483 dense core \cite{agundez2019:Sensitive}; 
all are cold sources characterized with a rotational temperature of the order of 10~K. On ice-grain surfaces, \ce{CH3O} is an identified intermediate 
in the formation of complex organic molecules \cite{gutierrez-quintanilla2021:ICOM} and a central intermediate in the hydrogenation scheme of 
CO toward methanol \cite{brown1988:Model, tielens1997:Ices, watanabe2002:Efficient}. 
Its presence in the gas-phase can result from non-thermal desorption from ice-grains or chemistry in the gas phase, 
possibly photodissociation of methanol or formaldehyde \cite{cernicharo2012:CH3O, bacmann2016:Origin}, ion-molecule reactions 
\cite{bacmann2016:Origin}, and methanol oxidation with OH \cite{cernicharo2012:CH3O, antinolo2016:Reactivity}. 

In this context, special interest has been given to the relationship between \ce{CH3O} and its \ce{CH2OH} isomer (of lower energy) 
as the predominance of one or the other radical in the gas phase could reflect desorption from ice grains or gas phase chemistry 
(see, e.g., Ref.\cite{bermudez2017:Laboratory} and Refs. within).  Their formation on ice-grain surface upon UV irradiation 
of interstellar ice analogs has been monitored \cite{gerakines1996, watanabe2002:Efficient, watanabe2007:Laboratory, chuang2016:Hatom} 
and their isomerization has also been subject to significant theoretical attention 
\cite{tachikawa1993:Reaction, cui1996:Initio, wang2010:Radical, burgers2005:Acidcatalyzed, wang2012:Hydrogen}; 
the latter works suggesting that both isomers should be stable in the interstellar medium. 
Interestingly, while rotational frequencies for the \ce{CH2OH} radical have now been available for a few years 
\cite{bermudez2017:Laboratory, chitarra2020:Reinvestigation, coudert2022:CH2OH}, the \ce{CH2OH} radical is yet to be detected 
in the interstellar medium.
While radicals, such as methoxy, are usually not found in dense and warm environments of star-forming regions, 
they are frequently detected in luke-warm ($\sim$50~K) sources, for example in photo-dissociation regions, 
such as the Orion Bar \cite{Orion-Bar_mols_w_2-4C_2015,Orion-Bar_COMs_2017}.

In this work, we report an extension of the observation of the pure rotational spectrum of \ce{CH3O} at submillimeter wavelengths up to almost 900~GHz. The newly measured transitions were fit together with literature pure rotational data to produce the most reliable set of spectroscopic parameters for the radical to date. These parameters and corresponding spectral predictions can be used with confidence for future searches for the radical, notably in luke-warm to warm interstellar sources.

\section{Experimental method}
The room-temperature pure rotational spectrum of the \ce{CH3O} radical has been recorded between 350 and 860~GHz using a source-frequency modulated submillimeter-wave spectrometer exploited in a double-modulation configuration, as described in  \citet{chitarra2022:CH2CN}. The \ce{CH3O} radical was produced by H-abstraction from methanol using experimental conditions described in our previous studies on the \ce{CH2OH} radical, the second main product of the reaction \cite{coudert2022:CH2OH}.
In brief, a constant flow of methanol was injected through a reaction cell where it encountered a flow of F-atoms produced by microwave discharge in a 5\% \ce{F2}:He mixture and injected using three inlets, with partial pressures of 10~\textmu bar (1~Pa) for methanol and 20~\textmu bar (2~Pa) for the F-mixture.
The content of the reaction cell was probed in absorption by the submillimeter-wave radiation emitted by a frequency multiplication chain. The frequency of this radiation was modulated (at $f=48$~kHz) with modulation depths adjusted to the probed spectral regions (from 0.5 to 1.8~MHz). A double-pass configuration was achieved using a roof-top mirror rotated by 45\textdegree\ from the vertical axis, resulting in a reflected beam with a polarization rotated by 90\textdegree\ from the incident radiation.  A polarizing grid placed at the entrance of the cell directed the reflected radiation towards an off-axis parabolic mirror focusing the radiation onto the detector (either a Schottky diode or a liquid-helium cooled InSb hot electron bolometer).
A current, generated by a waveform generator followed by an audio amplifier and a diode, circulating in a solenoid placed around the reaction cell allowed to generate an alternating magnetic field (at $f'=71$~Hz). Using two lock-in amplifiers, the submillimeter-wave signal detected was first demodulated at the second harmonic of the source frequency modulation and then demodulated at the frequency of the magnetic field modulation. This double-modulation configuration ensured that only the spectral lines of paramagnetic species are visible in the resulting spectrum (with some limitations that will be discussed in section \ref{sec:DM}).

\section{Results and Discussion}

\subsection{Spectroscopic assignments}

\begin{figure}[ht!]
    \centering
    \includegraphics[width=\columnwidth]{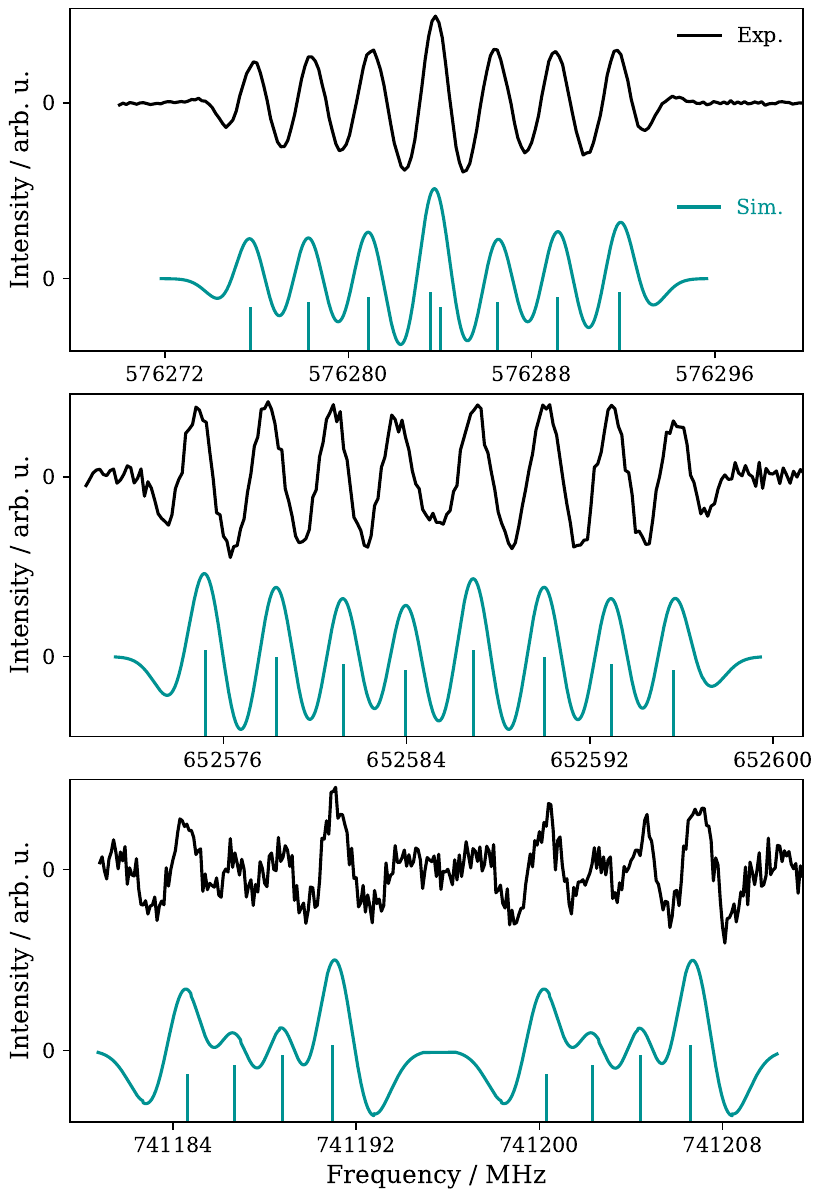}
    \caption{Evolution of the hyperfine splittings in the $N'-N''=10-9$, $12-11$, and $13-2$, $|K|=2, \Lambda = 1$ transitions of \ce{CH3O} (from top to bottom). The experimental spectrum (double modulation, in black) is compared to a simulation of \ce{CH3O} spectrum (final fit, in blue green). The stick spectra correspond to the catalog list (with intensities in linear units) and the full trace is a convolution with the second derivative of a Gaussian profile (full-width-at-half-maximum set to 1.3 times the expected Doppler value). }
    \label{fig:spectro}
\end{figure}

Experimental searches and spectroscopic assignments were based on the catalog entry found in the JPL database \cite{JPL-catalog_1998}. Many transitions display partially to totally-resolved hyperfine structure as shown in Figure~\ref{fig:spectro}. 
The signal-to-noise ratio (SNR) of the observed transitions is excellent up to about 600~GHz and decreases at higher frequencies, a reflection of the lower power emitted by the radiation source and higher absorption by the teflon optics. The observed transitions of \ce{CH3O} display a full-width-at-half-maximum (FWHM) slightly higher than what expected from Doppler considerations alone (about 1.3 times the expected value, Figure~\ref{fig:spectro}). This results from a combination of i) a modulation depth of the frequency modulation slightly larger than the natural width of the lines which results in an enhanced SNR at the expense of the FWHM; and ii) a broadening of the lines by Zeeman effect (either or both Earth-based and induced by the solenoid).

In total, 518 experimental transitions with 321 different frequencies involving $N$ and $K$ values up to 15 and 7, respectively, could be assigned with confidence between 351947 and 853920~MHz. Of these, 102 lines (62 frequencies) between 351 and 372~GHz were previously reported in \citet{momose1988:Submillimeterwave}.

\subsection{Hamiltonian}
The rotational spectrum of methoxy was calculated and fit applying the SPCAT and SPFIT programs \cite{spfit_1991}, respectively.
The programs use Hund's case (b) quantum numbers and parameters. Even though methoxy is closer to Hund's case (a) at lower quantum numbers, 
earlier work has shown that this is not a drawback to model spectra of such a radical. 
We mention, for example, studies on OH \cite{OH_SH_rot_2012,drouin2013:OH},  NO \cite{NO_rot_2015,NO_fitting_2017}, and IO \cite{IO_rot_2001} 
which adopted SPFIT to analyze the spectra.

The effective Hamiltonian used to model the data is similar to that used by \citet{endo1984:Microwavespectrum} as implemented in the 
\ce{CH3O} entry of the JPL catalog \footnote{\url{https://spec.jpl.nasa.gov/ftp/pub/catalog/archive/}, entry \texttt{031006}, visited on 2024/12/17} 
\cite{JPL-catalog_1998}. It can be expressed as:
\begin{equation}
    \mathcal{H} = \mathcal{H}_\mathrm{SO+COR} + \mathcal{H}_\mathrm{ROT+CD} + \mathcal{H}_\mathrm{SR+CD} + \mathcal{H}_\mathrm{HFS}
\end{equation}
where $\mathcal{H}_\mathrm{SO+COR}$ describes the electron spin-orbit and electronic Coriolis interaction, $\mathcal{H}_\mathrm{ROT+CD}$ the rotational energy and centrifugal distortion effect, $\mathcal{H}_\mathrm{SR+CD}$ the electron spin-rotation interaction with centrifugal distortion correction, and  $\mathcal{H}_\mathrm{HFS}$ the hyperfine interaction. We refer the reader to \citet{endo1984:Microwavespectrum} and \citet{momose1988:Submillimeterwave} for detailed descriptions of each term.
The initial set of spectroscopic parameters was based on the one from the archive of the JPL catalog even though the catalog entry was only based on \citet{endo1984:Microwavespectrum}. 
While it is recommended to use $A - B$ and $B$ to fit a symmetric top molecule with SPFIT, it is also possible to fit $A$ and $B$ if one uses the asymmetric rotor codes for $B$ and $C$ and locks $C$ to being equal to $B$. Both approaches yield the same results (parameters value and error) in the present work.

Trial fits revealed that the latest version of SPFIT\footnote{\url{https://spec.jpl.nasa.gov/ftp/pub/calpgm/whats.new}, visited on 2024/12/17}, 
visited on 2024/12/17 (from 2007) rejected a small number of lines. Trial calculations with SPCAT showed that one of the levels 
of such transition was severely displaced from the others and had a mixing coefficient of exactly zero. 
It was necessary to fit and calculate these levels properly as these affect some low-energy transitions which are very important in astronomical observations. 
It turned out that a version from 2006 treated these lines properly, which does not surprise given that the catalog entry is from early 2006.

As a consequence of using the 2006 version of SPFIT, a maximum of 6 quantum numbers per state can be used. Each rotational level is thus described using 5 quantum numbers, namely $N,K,\varv,t,F$, where $\varv=0$ for $\Lambda=1$ and $\varv=1$ for $\Lambda=0$, and $t$ is the spin designation and is a function of $N-F$, $J-F$, the symmetry (0, 2, 4 for A, E, E, respectively) and $I_{tot}$. The correspondence table between quantum numbers is available in the \texttt{d031006.pdf} file of the JPL entry.

\subsection{Fit of the transitions}
Besides the data from our present investigation, we used earlier published field-free pure rotational data \cite{endo1984:Microwavespectrum,momose1988:Submillimeterwave,laas2017:Millimeter}. 
Uncertainties of 30~kHz were assigned to the initial millimeter data \cite{endo1984:Microwavespectrum}, based on a statement in that work and commensurate with out final fit results. 
The subsequent submillimeter data \cite{momose1988:Submillimeterwave} were given uncertainties of 50~kHz, while the reported uncertainties were used for the most recent millimeter study \cite{laas2017:Millimeter}. 
The parity designations of the two earlier studies \cite{endo1984:Microwavespectrum,momose1988:Submillimeterwave} were reversed, following \citet{liu2009:CH3O}, where the study of the $^2\mathrm{A}_1 - {^2\mathrm{E}}$ electronic spectrum permitted to determine the parity unambiguously.
In the work of \citet{momose1988:Submillimeterwave}, several groups of transitions were not, or insufficiently, resolved as frequently observed for hyperfine splitting at high quantum numbers or at high frequencies. Because most fitting programs treat blended lines as separate pieces of information at the same frequency, the uncertainties reported by the authors were increased for these transitions. 
The SPFIT program permits to treat blended transitions at a certain frequency as intensity weighted average; equal intensity is assumed if no weighting is made. We adopted the normal uncertainties of 50~kHz for the small number of blended transitions separated by less than half of the assumed line width. 
In a very small number of cases, one transition of a group of three was omitted because it appeared to be too far away from the other, more closely spaced transitions to contribute substantially to the measured line frequency. 
In one case, an increased uncertainty was retained because of the fairly wide, partially resolved hyperfine pattern.

It is not easy to find the best set of spectroscopic parameters for such a complex spectroscopic problem with fine and hyperfine structure, Coriolis coupling, and extensive $l$-type resonance. 
In fact, it was pointed out in the early publications \cite{endo1984:Microwavespectrum,momose1988:Submillimeterwave,liu2009:CH3O} that many parameters were correlated and that it was difficult to find a set of spectroscopic parameters that is both meaningful and that reproduces the experimental lines to within experimental uncertainties on average. 
In the most recent study \cite{laas2017:Millimeter}, the $A$ rotational constant, first-order Coriolis coupling parameter $A\zeta _{t}$, and  $a\zeta _{e}d$---where $a$ is the spin-orbit coupling constant---were kept fixed to values from \citet{liu2009:CH3O}. These three parameters could be determined in the present work for reasons detailed in the following paragraphs.

It is useful to recall that the ground state rotational parameter $A_0$ ($C_0$) of a prolate (oblate) symmetric top as well as the purely $K$-dependent distortion parameters are in the general case not determinable by rotational or rovibrational spectroscopy. 
However, high-resolution spectroscopy of a prolate symmetric top between the ground and an excited state determine $\Delta A = A - A_0$ of that excited state directly or $A$, if $A_0$ is fixed to a certain value.  
Determining $\Delta K = 3$ loops are one way to overcome this problem. It requires the analyses of two degenerate vibrational fundamentals, their combination and hot bands. This was applied, for example, to \ce{CH3CN} \cite{MeCN_DeltaK=3_1993}. 
Utilizing forbidden transitions facilitated by rovibrational interactions are another means employed, used for example to determine $A$ and $D_K$ of \ce{CH3F} \cite{graner1976:CH3F} or to improve $A$, $D_K$, and $H_K$ of \ce{CH3CN} through rotational and rovibrational transitions \cite{MeCN_v8le2_2015,MeCN_up2v4eq1_etc_2021}. 
Another method is available if there are suitable near-degeneracies in $K$ within the ground vibrational state. This is a fairly common method in the case of oblate symmetric tops, such as \ce{PH3} \cite{PH3_analysis_2013}.
This method was also the one through which $A$ became determinable for \ce{CH3O} because of near-degeneracies with $\Delta l = \pm2$ and $\Delta K = \pm2$, $\Delta K = \mp1$, or $\Delta K = \mp4$ \cite{endo1984:Microwavespectrum,momose1988:Submillimeterwave,liu2009:CH3O}.

The Coriolis parameter $A\zeta$ within a doubly degenerate vibrational state is usually determinable if $A$ was determined or is fixed to a certain value. It was thus possible to adjust it in the present work.
Finally, the spin-orbit splitting in a linear molecule with, for instance, a $^2\Pi$ ground state is difficult to determine through rotational spectroscopy if the rotational energy value of the accessed transitions are much smaller than the spin-orbit splitting, but it becomes well determinable if the rotational energy approaches or even exceeds this value, again something that was made possible by the present investigations.

In order to determine a parameter set that is as large as necessary, but about as small as possible and fairly well constrained, we checked the parameter set for very high correlations and searched for parameters to add that are well determined, meaning their values are substantial multiples of their uncertainties, and that their inclusion improves the rms error, also known as weighted rms, by a sufficient amount. 
We also checked frequently among the parameters with fairly large uncertainties if they may be omitted without increasing the rms error too much. 
The initial parameter set contained $A$, $A\zeta _{t}$, $\eta _{K}$, $a\zeta _{e}d$, $\epsilon _{aa}$, $D^{s}_{K}$, and a distortion correction called $a_D\zeta _{e}d$, but coded as a $K$ correction to $a\zeta _{e}d$ instead of a $N$ correction. 
The fit was very unstable, meaning very many iterations were necessary to reach convergence, and many of the parameters were correlated. Almost perfect correlation occurred between $a\zeta _{e}d$ and its distortion correction; consequently, the latter parameter was omitted. 
The omission of a parameter so large in magnitude compared to its parent parameter $a\zeta _{e}d$ had a substantial impact on many parameters. Remarkably, the values of $D^{s}_{K}$ and $D^{s}_{KN}$ changed greatly in magnitude, such that both could be omitted without substantial further increase of the rms error.

The great number of transition frequencies determined with microwave accuracy that was added in the present study to the existing line list, and in particular the increase in frequency and quantum numbers, required several distortion parameters to be added with respect to previous parameter sets. 
The inclusion of $H_{KN}$, which was employed in some fits earlier \cite{endo1984:Microwavespectrum,momose1988:Submillimeterwave}, and two octic centrifugal distortion parameters was obvious; $H_{NK}$ was retained in the fit despite being not determined significantly because it is a lower order relative of $L_{NNK}$, a parameter that significantly reduced the rms of the fit. 
The parameter $D_{K}$ is on the verge of being determinable, its uncertainty was about 1~MHz in test fits, comparable to its magnitude. Therefore, it was kept fixed to the value determined for \ce{CH3F} \cite{graner1976:CH3F}, as done in earlier publications. 
We added two quartic Coriolis distortion parameters $\eta _{NK}$ and $\eta _{NN}$. 
%
With respect to the fine structure parameters, we applied $a_N\zeta _{e}d$ in the fit, as in most earlier studies. The inclusion of this parameter permitted to omit the $l$-type resonance parameter $h_{2N}$ whose value in the fit prior to trying out $a_N\zeta _{e}d$ was $-$0.46~kHz $\pm$ 0.08~kHz. 
This distortion parameter is slightly better determined than in previous studies and allows to reduce the rms error of the fit slightly more than $h_{2N}$. 
As mentioned above, $D^{s}_{K}$ and $D^{s}_{KN}$ were omitted from the fit. $\epsilon _{4}$ was included temporarily but omitted in the final fit.
Distortion corrections to the hyperfine structure parameters, nuclear spin-rotation, or nuclear spin-nuclear spin coupling parameters had negligible effects in the fit and were not kept in the final fit.

It is surprising that the $\Delta K = 3$ parameters $\alpha$ or $\beta$ were not considered in earlier analyses. They are the coefficients of $[N_+^3 + N_-^3, N_z]$ and of $i( N_+^3 - N_-^3)$, respectively, with $N_{\pm} = N_x \pm N_y$ and [A,B] = AB + BA being the anticommutator. 
We mention that $\alpha$ is frequently called $\epsilon$; we do not use the latter designation here because of possible confusion with the spin-rotation parameters. 
The parameter $\alpha$ is important, for example, to describe spectra of oblate symmetric tops, such as \ce{PH3} \cite{PH3_analysis_2013}.
Trial fits with $\beta$ included reduced the rms error somewhat, produced a reasonably well determined parameter, but the convergence of the fit was very slow, such that this parameter was discarded. 
Fits with $\alpha$ included reduced the rms error similarly well as $\beta$, but the convergence was very fast. Moreover, the initially fairly well determined $l$-type resonance parameter $h_4$ (194~Hz $\pm$ 11~Hz) could be omitted as well as the intermediately introduced $\epsilon _{1N}$ and $\epsilon _4$, which were moderately and fairly well determined, respectively, with only a small increase in the rms error. 
Several other parameters were tested, including the spin-rotation parameters related to $\alpha$ and $\beta$. None of them was retained as none impacted the fit sufficiently.

The final spectroscopic parameters are gathered in Table~\ref{methoxy-spec-parameters} together with the ones from the most recent previous study \cite{laas2017:Millimeter}. 
Selected parameter values from earlier studies are mentioned in the discussion.
The final line list consists of 970 transitions corresponding to 626 different frequencies due to unresolved hyperfine or asymmetry splitting. Because 29 transitions (20 frequencies) from \citet{laas2017:Millimeter} and 102 transitions (62 frequencies) from this work were already reported in \citet{momose1988:Submillimeterwave}, the final dataset consists of 839 transitions observed in 544 individual lines. The overall rms error of the fit is 0.961, thus the transition frequencies were fit to the assigned uncertainties on average. 
In detail, the 126 transitions of \citet{endo1984:Microwavespectrum} (102 frequencies, $N \leq 4$, $|K| \leq 3$) were fit to 29.8~kHz, which corresponds to an rms error of 0.992. 
The 291 transitions from \citet{momose1988:Submillimeterwave} (179 frequencies, $N \leq 7$, $|K| \leq 6$) were fit to 53.4~kHz, thus with an rms error of 1.059. 
The 67 transitions by \citet{laas2017:Millimeter} (48 frequencies, $N \leq 7$, $|K| \leq 3$) were fit to an rms error of 0.938 or 66.5~kHz. 
Our 518 transitions (321 frequencies, $N \leq 15$, $|K| \leq 7$) were fit to an rms error of 0.895 or 128.2~kHz.

The fit files (list of transitions, parameters, and fit output) are available as supplementary material to this article. 
The line, parameter, and fit files along with auxiliary files are also available in the fitting spectra 
section\footnote{\url{https://cdms.astro.uni-koeln.de/classic/predictions/pickett/beispiele/CH3O/}, visited on 2024/12/17} 
of the Cologne Database for Molecular Spectroscopy, CDMS, \cite{CDMS_2001,CDMS_2005,CDMS_2016}. 
The auxiliary files include information on the partition function and properties of the molecule and the calculation.
A line list of the rotational spectrum of methoxy at 300~K is available in the entries 
section\footnote{\url{https://cdms.astro.uni-koeln.de/classic/entries/}, visited on 2024/12/17} of the CDMS. 
The permanent dipole moment of 2.12~D, employed for this calculation, is from a quantum chemical calculation \cite{CH3O_ai_dip_etc_1982}.


\begin{table}[ht!]
  \begin{center}
  \caption{Spectroscopic parameters$^a$ (MHz) of the methoxy radical from the present work 
           in comparison to those from the latest previous work$^b$.}
  \label{methoxy-spec-parameters}
\smallskip
{\footnotesize
\renewcommand{\arraystretch}{1.10}
  \begin{tabular}{lr@{}lr@{}l}
  \hline
Parameter                                           &\multicolumn{2}{c}{Present} &\multicolumn{2}{c}{Previous}\\[1pt]
\hline
$A$                                                 &      155491&.~(36)         &      154670&.$^c$         \\
$B$                                                 &       27930&.10487~(76)    &       27930&.101~(5)      \\
$D_{K}$                                             &           2&.1084$^d$      &           2&.1084$^d$     \\
$D_{NK} \times 10^3$                                &         746&.289~(194)     &         768&.2~(2)        \\
$D_{N} \times 10^3$                                 &          75&.4172~(44)     &          75&.26~(5)       \\
$H_{KN} \times 10^6$                                &      $-$139&.4~(114)       &            &              \\
$H_{NK} \times 10^6$                                &           2&.47~(162)      &            &              \\
$L_{KKN} \times 10^6$                               &           2&.942~(199)     &            &              \\
$L_{NNK} \times 10^9$                               &          52&.3~(52)        &            &              \\
$\alpha \times 10^3$                                &       $-$21&.59~(62)$^e$   &            &              \\
$A\zeta _{t}$                                       &       51348&.4~(45)        &       52036&.$^c$         \\
$\eta _{K}$                                         &       $-$23&.46~(82)       &       $-$16&.1~(3)        \\
$\eta _{N} \times 10^3$                             &      $-$168&.77~(28)       &      $-$160&.~(1)         \\
$\eta _{NK} \times 10^3$                            &          13&.343~(74)      &            &              \\
$\eta _{NN} \times 10^6$                            &         375&.88~(186)      &            &              \\
$h_{1}$                                             &       $-$75&.1383~(92)     &       $-$75&.346~(68)     \\
$h_{1K} \times 10^3$                                &      $-$190&.7~(74)        &         116&.~(67)        \\
$h_{1N} \times 10^3$                                &           1&.3680~(158)    &           1&.297~(174)    \\
$h_{2}$                                             &     $-$1298&.861~(77)$^e$  &     $-$1309&.03~(5)       \\
$h_{2K} \times 10^3$                                &      $-$578&.9~(54)$^e$    &          37&.~(3)         \\
$h_{2N} \times 10^3$                                &            &               &        $-$4&.0~(3)        \\
$h_{4} \times 10^6$                                 &            &               &      $-$633&.~(170)       \\
$a\zeta _{e}d$                                      &  $-$1841147&.~(186)        &  $-$1843572&.$^c$         \\
$a_N\zeta _{e}d$                                    &        $-$0&.778~(123)     &            &              \\
$\epsilon _{aa}$                                    &    $-$40306&.1~(68)        &    $-$37487&.~(73)        \\
$\epsilon _{bb}$                                    &     $-$1052&.2~(62)        &     $-$1114&.7~(3)        \\
$\epsilon _{1}$                                     &         172&.393~(12)      &         172&.592~(46)     \\
$\epsilon _{2}$$^e$                                 &           2&.20~(30)       &            &              \\
$D^{s}_{K}$                                         &            &               &       $-$16&.9~(3)        \\
$D^{s}_{KN}$                                        &            &               &        $-$1&.04~(12)      \\
$D^{s}_{NK}$                                        &        $-$1&.7928~(44)     &        $-$0&.336~(34)     \\
$D^{s}_{N}$                                         &        $-$0&.03528~(56)    &        $-$0&.0312~(40)    \\
$a_{L}$                                             &           2&.3344~(100)    &           2&.264~(52)     \\
$\sigma_{0}$                                        &         120&.427~(74)      &         120&.55~(35)      \\
$\sigma_{\pm}$                                      &         153&.00~(31)       &         152&.22~(155)     \\
$T_{0}^{2}(C_{0})$                                  &           5&.443~(54)      &           5&.380~(270)    \\
$T_{0}^{2}(C_{\pm})$                                &          54&.95~(31)       &          55&.17~(159)     \\
$T_{\pm2}^{2}(C_{0})$                               &        $-$0&.3402~(85)     &        $-$0&.349~(47)     \\
$T_{\mp2}^{2}(C_{\pm})$                             &           1&.799~(31)      &           1&.791~(96)     \\
\hline
\end{tabular}\\[2pt]
}
\end{center}
$^a$ 
Numbers in parentheses are 1~$\sigma$ uncertainties in units of the least significant figures. 
Parameters without uncertainties were kept fixed in the analysis. Empty fields indicate 
that this parameter was not used in the fit.\\
$^b$ 
Ref.~\cite{laas2017:Millimeter}.\\ 
$^c$ 
Kept fixed to value from Ref. \cite{liu2009:CH3O}.\\
$^d$ 
Kept fixed to value from \ce{CH3F} \cite{graner1976:CH3F}.\\
$^e$
The relative signs of $\alpha$, $h_{2}$, $h_{2K}$, and $\epsilon _{2}$ are determined; they may be inverted simultaneously.\\
\end{table}


Our extended line list, with its significant increase in observed $N$ values (15 compared to 7 previously), constrains the rotational spectrum of methoxy, in particular to higher frequencies and higher quantum numbers than those observed in the literature and included in previous lists. 
The comparison of the present spectroscopic parameters with those from the latest previous study \cite{laas2017:Millimeter} in Table~\ref{methoxy-spec-parameters} reveals aspects that may surprise at first sight, but are in line with the extensive new measurements and fit evolution. 
We believe that some of the previously reported parameters were taking values resulting from error compensations---a reflection of the limited dataset---and that the present extensive measurements allow to lift these ambiguities, yielding a more meaningful set of parameters.
The values of $B$, $D_{N}$, $h_{1}$, $h_{1N}$, $h_{2}$, and the hyperfine parameters are very similar to those of Ref.  \cite{endo1984:Microwavespectrum, momose1988:Submillimeterwave, liu2009:CH3O, laas2017:Millimeter} as far as they were determined.
There are, however, considerable changes in the highly $K$-dependent parameters, even those of low order, often well outside the uncertainties, while the more $N$-dependent parameters have more similar values. 
No hyperfine splitting was considered in Ref.~\cite{liu2009:CH3O}. 
Two hyperfine parameters in Ref.~\cite{ momose1988:Submillimeterwave} are smaller than ours by a factor of $\sim$1.5 but we attribute this to a different definition of these parameters in that study.

The large changes in $A$, $A\zeta _{t}$, and $a\zeta _{e}d$ are less surprising, in light of the statements concerning their determinability above than it may seem at first sight. 
Fixing all three of these parameters in Ref.~\cite{laas2017:Millimeter} to values from Ref.~\cite{liu2009:CH3O} is questionable, in particular because $a_N\zeta _{e}d$, taking a large value in Ref.~\cite{liu2009:CH3O}, was omitted in Ref.~\cite{laas2017:Millimeter}. 
Uncertainties of $A$ and $a\zeta _{e}d$ are factors of 2.4 and 1.3, respectively, smaller than ours in the study of the electronic spectrum \cite{liu2009:CH3O} which probes $a\zeta _{e}d$ directly.  
The reduction in its uncertainty probably led in turn to a smaller uncertainty in $A$. 
This may offer the opportunity to improve the accuracies of these parameters slightly more by including the electronic data  from \citet{liu2009:CH3O} work into future fit.

The parameter $\alpha$ was determined for methoxy for the first time in our study. Its magnitude of $\sim$22~kHz may be compared with 96~kHz derived for \ce{CH3D} \cite{CH3D_v0_1999} or 715~kHz determined for \ce{PH3} \cite{PH3_analysis_2013}.

It is worthwhile mentioning that some of the present higher order parameters are much smaller in magnitude than values from earlier studies. We discuss in the following only values determined with sufficient significance.
Values of $(363 \pm 25)$~MHz and $(193 \pm 6)$~MHz were determined for $a_N\zeta _{e}d$ in Ref. \cite{ momose1988:Submillimeterwave} and Ref. \cite{liu2009:CH3O}, respectively, whereas our value is $(-0.778 \pm 0.123)$~MHz. 
The uncertainty of $a_K\zeta _{e}d$ in a trial fit was 4.4~MHz with its value much smaller in magnitude.
Similarly, a value of $(-679 \pm 79)$~MHz was determined for $\epsilon _{2a}$ \cite{liu2009:CH3O}; our value for $\epsilon _{2} = (\epsilon _{2a} + \epsilon _{2b})/2$ is $(2.20 \pm 0.30)$~MHz. 
There are other cases with usually somewhat smaller differences.

\subsection{Astronomical implications}
\begin{figure}[ht!]
    \centering
    \includegraphics[width=\columnwidth]{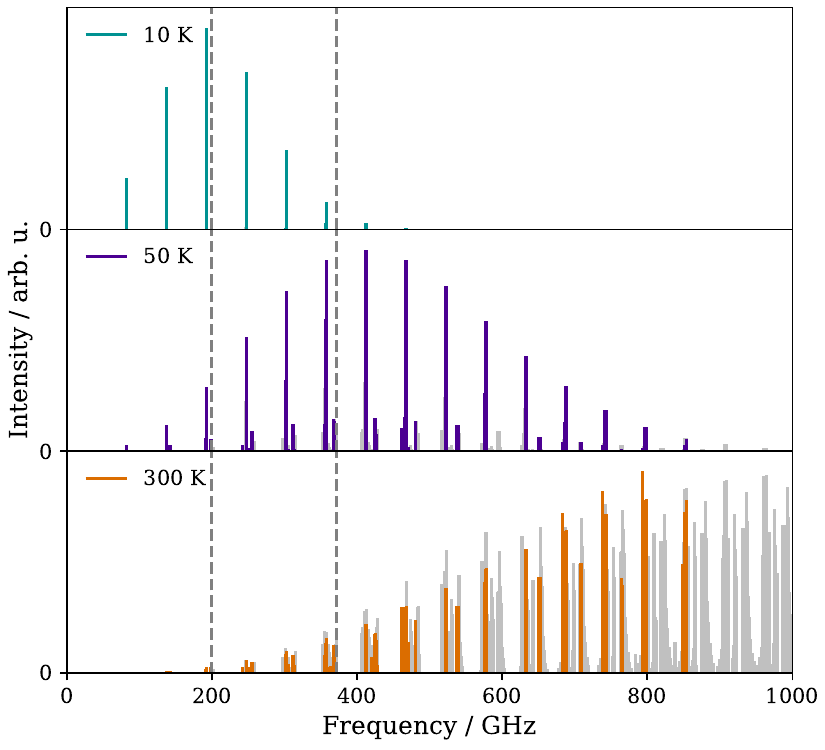}
    \caption{Simulations of the pure rotational spectrum of \ce{CH3O} in $\varv=0$ up to 1~THz at 10~K (in blue-green), 50~K (in purple) and 300~K (in orange). Lines measured in the literature and in this study are plotted in color while those that remain to be measured are in gray. The dashed vertical lines indicate the highest frequencies from the spectral catalog available on JPL (199~GHz) and in the previous literature (372~GHz).}
    \label{fig:cat}
\end{figure}

As mentioned already previously, the present linelist for \ce{CH3O} is much more reliable than previous ones to search for this species in the interstellar medium. It is particularly noteworthy that the previous catalog available in the JPL database was only based on the \citet{endo1984:Microwavespectrum} work (containing measured transitions below 199~GHz), while the present one contains all available pure rotational transitions of the radical reported in the literature \cite{endo1984:Microwavespectrum, momose1988:Submillimeterwave, laas2017:Millimeter} together with the present measurements. 
Figure~\ref{fig:cat} shows simulations of the spectrum of \ce{CH3O} below 1~THz at three temperatures (10, 50, 300~K, from top to bottom) with color coding highlighting the lines for which experimental frequencies have been determined and are included in our fit.
All observed transitions above 371~GHz are reported here for the first time. 

We bring to the reader's attention that recalculation of the spectral catalog (for instance at different temperatures or if new measured transitions are added to the fit) should be done using the 2006 version of SPCAT otherwise some transitions will not be predicted.

\subsection{Advantages and limitations of the double-modulation scheme}\label{sec:DM}
\begin{figure}[ht!]
    \centering
    \includegraphics[width=\columnwidth]{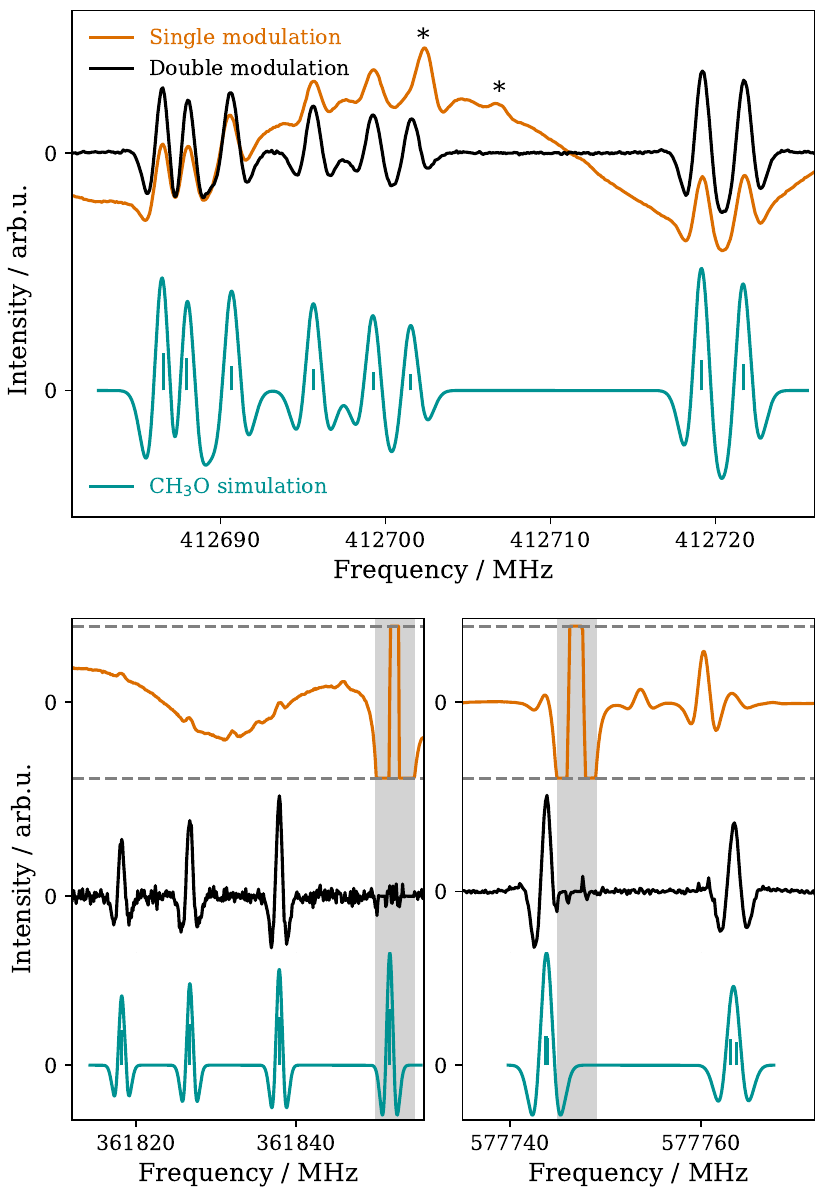}
    \caption{Comparison of spectra recorded in source-frequency modulation (single modulation, orange traces) and double-modulation (black traces) configurations with a simulation of \ce{CH3O} spectrum (final fit, in blue green). The stick spectra correspond to the catalog list (with intensities in linear units) and the full trace is a convolution with the second derivative of a Gaussian profile (full-width-at-half-maximum set to 1.3 times the expected Doppler value). [\textit{top}] Cluster of $N_{|K|}=7_3-6_3$ and $7_4-6_4, \Lambda=0$ lines of \ce{CH3O}. The lines identified by an asterisk belongs to a close-shell molecule. [\textit{bottom}] $N_{|K|}=7_5-6_5, \Lambda=1$ and $10_3-9_3, \Lambda=0$ lines of \ce{CH3O}. The saturation limit of the lock-in amplifier is indicated by the dashed horizontal lines and the resulting frequency portions of the spectrum where the signal cannot be exploited in double resonance configuration are shaded in gray.}
    \label{fig:DMlimits}
\end{figure}

The present study of \ce{CH3O} is a new illustration of both the strengths and weaknesses of double-modulation spectroscopy, as previously shown in \citet{chitarra2022:CH2CN} and \citet{coudert2022:CH2OH}. The main advantage of the technique is that lines of close-shell species are filtered out while those of radical species appear over a flat baseline (no residual Fabry-Perot effect) as shown in Figure~\ref{fig:DMlimits} (upper panel). This allows to unveil some transitions that would not be observable otherwise (for instance, the one close to the leftmost close-shell transition in Figure~\ref{fig:DMlimits}).
However, at the sensitivity level required to detect weak radical lines, very strong lines of close-shell molecules, typically those of the precursor, cause the signal to be saturated at the first demodulation stage. In that case, the signal is not exploitable at the second demodulation stage which can affect the line profile of some radical lines or even fully prevent their observation (Figure~\ref{fig:DMlimits}, gray areas in the bottom panels).
While this effect did not hinder the present work, it can completely impede the study of radical species produced from a precursor exhibiting a dense and intense rotational spectrum.

To overcome these limitations while still benefiting from radical selectivity, the experimental setup has been upgraded since the results reported in the present article. Currently, the set-up is exploited using either pure Zeeman modulation \cite{chahbazian2024:CH2CHO, martin-drumel2024:H2NCO} or Faraday rotation modulation \cite{chahbazian2024:Faraday}. We refer the reader to \citet{chahbazian2024:Faraday} for a detailed comparison of the performances of the two configurations.

\section{Conclusion}

Extensive measurements of \ce{CH3O} have been performed over the entire sub-millimeter spectral region allowing to significantly extend the range of observed frequencies and quantum numbers of the radical in its vibrational ground state.
The newly measured data have been combined with pure rotational transitions measured by \citet{endo1984:Microwavespectrum, momose1988:Submillimeterwave, laas2017:Millimeter} and fit with an effective Hamiltonian using SPFIT/SPCAT.
The spectroscopic parameters are determined with confidence which allows for reliable prediction up to about 1~THz. However, we recall to the reader's attention that uncertainties increase rapidly, in particular for the weaker transitions. All spectral assignments, fit, and prediction files are made available as supplementary material and in the CDMS database.

\section*{Acknowledgements}
The work at ISMO was performed thanks to financial support from LabEx PALM (ANR-10-LABX-0039-PALM), from the Région Ile-de-France through DIM-ACAV+, and from the Agence Nationale de la Recherche (ANR-19-CE30-0017-01). We acknowledge support from the Programme National ``Physique et Chimie du Milieu Interstellaire'' (PCMI) of CNRS/INSU with INC/INP co-funded by CEA and
CNES. We are also grateful to the AILES beamline staff from the SOLEIL synchrotron for the loan of the bolometer.
H.S.P.M. acknowledges support by the Deutsche Forschungsgemeinschaft via the collaborative research 
center SFB~1601 (project ID 500700252) subprojects A4 and Inf.


\appendix


\bibliographystyle{elsarticle-num-names} 
\biboptions{sort&compress}
\bibliography{biblio}

\begin{thebibliography}{69}
\expandafter\ifx\csname natexlab\endcsname\relax\def\natexlab#1{#1}\fi
\providecommand{\url}[1]{\texttt{#1}}
\providecommand{\href}[2]{#2}
\providecommand{\path}[1]{#1}
\providecommand{\DOIprefix}{doi:}
\providecommand{\ArXivprefix}{arXiv:}
\providecommand{\URLprefix}{URL: }
\providecommand{\Pubmedprefix}{pmid:}
\providecommand{\doi}[1]{\href{http://dx.doi.org/#1}{\path{#1}}}
\providecommand{\Pubmed}[1]{\href{pmid:#1}{\path{#1}}}
\providecommand{\bibinfo}[2]{#2}
\ifx\xfnm\relax \def\xfnm[#1]{\unskip,\space#1}\fi
\bibitem[{Marenich and Boggs(2005)}]{marenich2005:Model}
\bibinfo{author}{A.~V. Marenich}, \bibinfo{author}{J.~E. Boggs},
\newblock \bibinfo{title}{A model spin-vibronic {{Hamiltonian}} for twofold
  degenerate electron systems: {{A}} variational {\emph{ab initio}} study of
  {{{\~X}}} {\textsuperscript{2}}{{E}} {{CH}}{\textsubscript{3}}{{O}}},
\newblock \bibinfo{journal}{J. Chem. Phys.} \bibinfo{volume}{122}
  (\bibinfo{year}{2005}) \bibinfo{pages}{024308}.
  \DOIprefix\doi{10.1063/1.1824878}.
\bibitem[{Shao and Mo(2013)}]{shao2013:JahnTeller}
\bibinfo{author}{Z.~Shao}, \bibinfo{author}{Y.~Mo},
\newblock \bibinfo{title}{Jahn-{{Teller}} effect in
  {{CH}}{\textsubscript{2}}{{DO}}/{{CHD}}{\textsubscript{2}}{{O}} ({{X}}
  {\textsuperscript{2}}{{E}}): {{Vibronic}} coupling of all vibrational modes},
\newblock \bibinfo{journal}{J. Chem. Phys.} \bibinfo{volume}{138}
  (\bibinfo{year}{2013}) \bibinfo{pages}{244309}.
  \DOIprefix\doi{10.1063/1.4811517}.
\bibitem[{Yarkony et~al.(1974)Yarkony, Schaefer, and
  Rothenberg}]{yarkony1974:Geometries}
\bibinfo{author}{D.~R. Yarkony}, \bibinfo{author}{H.~F. Schaefer},
  \bibinfo{author}{S.~Rothenberg},
\newblock \bibinfo{title}{Geometries of methoxy radical ({{X}}
  {\textsuperscript{2}}{{E}} and {{A}}
  {\textsuperscript{2}}{{A}}{\textsubscript{1}} states) and methoxide ion},
\newblock \bibinfo{journal}{J. Am. Chem. Soc.} \bibinfo{volume}{96}
  (\bibinfo{year}{1974}) \bibinfo{pages}{656--659}.
  \DOIprefix\doi{10.1021/ja00810a003}.
\bibitem[{Bersuker(2001)}]{bersuker2001:Modern}
\bibinfo{author}{I.~B. Bersuker},
\newblock \bibinfo{title}{Modern aspects of the {{Jahn}}-{{Teller effect
  theory}} and {{applications to molecular problems}}},
\newblock \bibinfo{journal}{Chem. Rev.} \bibinfo{volume}{101}
  (\bibinfo{year}{2001}) \bibinfo{pages}{1067--1114}.
  \DOIprefix\doi{10.1021/cr0004411}.
\bibitem[{Höper et~al.(2000)Höper, Botschwina, and
  Köppel}]{hoeper2000:Theoretical}
\bibinfo{author}{U.~Höper}, \bibinfo{author}{P.~Botschwina},
  \bibinfo{author}{H.~Köppel},
\newblock \bibinfo{title}{Theoretical study of the {{Jahn-Teller}} effect in
  {{{\~X}}} {\textsuperscript{2}}{{E CH}}{\textsubscript{3}}{{O}}},
\newblock \bibinfo{journal}{J. Chem. Phys.} \bibinfo{volume}{112}
  (\bibinfo{year}{2000}) \bibinfo{pages}{4132--4142}.
  \DOIprefix\doi{10.1063/1.480998}.
\bibitem[{Nagesh et~al.(2014)Nagesh, Sibert, and
  Stanton}]{nagesh2014:Simulation}
\bibinfo{author}{J.~Nagesh}, \bibinfo{author}{E.~L. Sibert},
  \bibinfo{author}{J.~F. Stanton},
\newblock \bibinfo{title}{Simulation of laser excitation spectrum of
  {{CH}}{\textsubscript{3}}{{O}} and {{CD}}{\textsubscript{3}}{{O}}},
\newblock \bibinfo{journal}{Spectrochim. Acta A} \bibinfo{volume}{119}
  (\bibinfo{year}{2014}) \bibinfo{pages}{90--99}.
  \DOIprefix\doi{10.1016/j.saa.2013.02.037}.
\bibitem[{Marenich and Boggs(2005)}]{marenich2005:Role}
\bibinfo{author}{A.~V. Marenich}, \bibinfo{author}{J.~E. Boggs},
\newblock \bibinfo{title}{The role of the cubic and quartic {{Jahn-Teller}}
  coupling in the {{{\~X}}} {\textsuperscript{2}}{{E}} ground electronic state
  of the methoxy radical {{CH}}{\textsubscript{3}}{{O}}},
\newblock \bibinfo{journal}{Chem. Phys. Lett.} \bibinfo{volume}{404}
  (\bibinfo{year}{2005}) \bibinfo{pages}{351--355}.
  \DOIprefix\doi{10.1016/j.cplett.2005.01.116}.
\bibitem[{El~Hilali et~al.(2009)El~Hilali, Boudon, and
  Loete}]{elhilali2009:Tensorial}
\bibinfo{author}{A.~El~Hilali}, \bibinfo{author}{V.~Boudon},
  \bibinfo{author}{M.~Loete},
\newblock \bibinfo{title}{Tensorial development of the rovibronic
  {{Hamiltonian}} and dipole moment operators for {{XY}}(3){{Z}} molecules with
  a degenerate electronic state: {{Preliminary}} application to the
  {{CH}}{\textsubscript{3}}{{O}} radical},
\newblock \bibinfo{journal}{J. Mol. Spectrosc.} \bibinfo{volume}{253}
  (\bibinfo{year}{2009}) \bibinfo{pages}{92--98}.
  \DOIprefix\doi{10.1016/j.jms.2008.10.007}.
\bibitem[{Radford and Russell(1977)}]{radford1977:Spectroscopic}
\bibinfo{author}{H.~E. Radford}, \bibinfo{author}{D.~K. Russell},
\newblock \bibinfo{title}{Spectroscopic detection of methoxy
  ({{CH}}{\textsubscript{3}}{{O}})},
\newblock \bibinfo{journal}{J. Chem. Phys.} \bibinfo{volume}{66}
  (\bibinfo{year}{1977}) \bibinfo{pages}{2222--2224}.
  \DOIprefix\doi{10.1063/1.434142}.
\bibitem[{Russell and Radford(1980)}]{russell1980:Analysis}
\bibinfo{author}{D.~K. Russell}, \bibinfo{author}{H.~E. Radford},
\newblock \bibinfo{title}{Analysis of the {{LMR}} spectra of methoxy,
  {{CH}}{\textsubscript{3}}{{O}}},
\newblock \bibinfo{journal}{J. Chem. Phys.} \bibinfo{volume}{72}
  (\bibinfo{year}{1980}) \bibinfo{pages}{2750--2759}.
  \DOIprefix\doi{10.1063/1.439423}.
\bibitem[{Ohbayashi et~al.(1977)Ohbayashi, Akimoto, and
  Tanaka}]{ohbayashi1977:Emissionspectra}
\bibinfo{author}{K.~Ohbayashi}, \bibinfo{author}{H.~Akimoto},
  \bibinfo{author}{I.~Tanaka},
\newblock \bibinfo{title}{Emission-spectra of {{CH}}{\textsubscript{3}}{{O}},
  {{C}}{\textsubscript{2}}{{H}}{\textsubscript{5}}{{O}}, and
  {{I-C}}{\textsubscript{3}}{{H}}{\textsubscript{7}}{{O}} radicals},
\newblock \bibinfo{journal}{J. Phys. Chem.} \bibinfo{volume}{81}
  (\bibinfo{year}{1977}) \bibinfo{pages}{798--802}.
  \DOIprefix\doi{10.1021/j100523a023}.
\bibitem[{Liu et~al.(2009)Liu, Chen, Melnik, Yi, and Miller}]{liu2009:CH3O}
\bibinfo{author}{J.~Liu}, \bibinfo{author}{M.-W. Chen},
  \bibinfo{author}{D.~Melnik}, \bibinfo{author}{J.~T. Yi},
  \bibinfo{author}{T.~A. Miller},
\newblock \bibinfo{title}{The spectroscopic characterization of the methoxy
  radical. {{I}}. {{Rotationally}} resolved {{A}}
  {\textsuperscript{2}}{{A}}{\textsubscript{1}} $-$ {{X}}
  {\textsuperscript{2}}{{E}} electronic spectra of
  {{CH}}{\textsubscript{3}}{{O}}},
\newblock \bibinfo{journal}{J. Chem. Phys.} \bibinfo{volume}{130}
  (\bibinfo{year}{2009}) \bibinfo{pages}{074302}.
  \DOIprefix\doi{10.1063/1.3072104}.
\bibitem[{Geers et~al.(1994)Geers, Kappert, Temps, and
  Wiebrecht}]{geers1994:Rotation}
\bibinfo{author}{A.~Geers}, \bibinfo{author}{J.~Kappert},
  \bibinfo{author}{F.~Temps}, \bibinfo{author}{J.~W. Wiebrecht},
\newblock \bibinfo{title}{Rotation--vibration state resolved unimolecular
  dynamics of highly vibrationally excited {{CH}}{\textsubscript{3}}{{O}}
  ({{X}} {\textsuperscript{2}}{{E}}). {{I}}. {{Observed}} stimulated emission
  pumping spectra},
\newblock \bibinfo{journal}{J. Chem. Phys.} \bibinfo{volume}{101}
  (\bibinfo{year}{1994}) \bibinfo{pages}{3618--3633}.
  \DOIprefix\doi{10.1063/1.467547}.
\bibitem[{Tsegaw et~al.(2016)Tsegaw, Sander, and Kaiser}]{tsegaw2016:Electron}
\bibinfo{author}{Y.~A. Tsegaw}, \bibinfo{author}{W.~Sander},
  \bibinfo{author}{R.~I. Kaiser},
\newblock \bibinfo{title}{Electron {{paramagnetic resonance spectroscopic
  study}} on {{nonequilibrium reaction pathways}} in the {{photolysis}} of
  {{solid nitromethane}} ({{CH}}{\textsubscript{3}}{{NO}}{\textsubscript{2}})
  and {{D3-nitromethane}}
  ({{CD}}{\textsubscript{3}}{{NO}}{\textsubscript{2}})},
\newblock \bibinfo{journal}{J. Phys. Chem. A} \bibinfo{volume}{120}
  (\bibinfo{year}{2016}) \bibinfo{pages}{1577--1587}.
  \DOIprefix\doi{10.1021/acs.jpca.5b12520}.
\bibitem[{Engelking et~al.(1978)Engelking, Ellison, and
  Lineberger}]{engelking1978:Laser}
\bibinfo{author}{P.~C. Engelking}, \bibinfo{author}{G.~B. Ellison},
  \bibinfo{author}{W.~C. Lineberger},
\newblock \bibinfo{title}{Laser photodetachment electron spectrometry of
  methoxide, deuteromethoxide, and thiomethoxide --- electron affinites and
  vibrational structure of {{CH}}{\textsubscript{3}}{{O}},
  {{CD}}{\textsubscript{3}}{{O}}, and {{CH}}{\textsubscript{3}}{{S}}},
\newblock \bibinfo{journal}{J. Chem. Phys.} \bibinfo{volume}{69}
  (\bibinfo{year}{1978}) \bibinfo{pages}{1826--1832}.
  \DOIprefix\doi{10.1063/1.436842}.
\bibitem[{Weichman et~al.(2017)Weichman, Cheng, Kim, Stanton, and
  Neumark}]{weichman2017:Lowlying}
\bibinfo{author}{M.~L. Weichman}, \bibinfo{author}{L.~Cheng},
  \bibinfo{author}{J.~B. Kim}, \bibinfo{author}{J.~F. Stanton},
  \bibinfo{author}{D.~M. Neumark},
\newblock \bibinfo{title}{Low-lying vibronic level structure of the ground
  state of the methoxy radical: {{Slow}} electron velocity-map imaging
  ({{SEVI}}) spectra and {{Koppel-Domcke-Cederbaum}} ({{KDC}}) vibronic
  {{Hamiltonian}} calculations},
\newblock \bibinfo{journal}{J. Chem. Phys.} \bibinfo{volume}{146}
  (\bibinfo{year}{2017}) \bibinfo{pages}{224309}.
  \DOIprefix\doi{10.1063/1.4984963}.
\bibitem[{Tang et~al.(2020)Tang, Lin, Garcia, Loison, Fittschen, Gu, Zhang, and
  Nahon}]{tang2020:Threshold}
\bibinfo{author}{X.~Tang}, \bibinfo{author}{X.~Lin}, \bibinfo{author}{G.~A.
  Garcia}, \bibinfo{author}{J.-C. Loison}, \bibinfo{author}{C.~Fittschen},
  \bibinfo{author}{X.~Gu}, \bibinfo{author}{W.~Zhang},
  \bibinfo{author}{L.~Nahon},
\newblock \bibinfo{title}{Threshold photoelectron spectroscopy of the methoxy
  radical},
\newblock \bibinfo{journal}{J. Chem. Phys.} \bibinfo{volume}{153}
  (\bibinfo{year}{2020}) \bibinfo{pages}{031101}.
  \DOIprefix\doi{10.1063/5.0016146}.
\bibitem[{Han et~al.(2002)Han, Utkin, Chen, Burns, and
  Curl}]{han2002:Highresolution}
\bibinfo{author}{J.~X. Han}, \bibinfo{author}{Y.~G. Utkin},
  \bibinfo{author}{H.~B. Chen}, \bibinfo{author}{L.~A. Burns},
  \bibinfo{author}{R.~F. Curl},
\newblock \bibinfo{title}{High-resolution infrared spectra of the {{C-H}}
  asymmetric stretch vibration of jet-cooled methoxy radical
  ({{CH}}{\textsubscript{3}}{{O}})},
\newblock \bibinfo{journal}{J. Chem. Phys.} \bibinfo{volume}{117}
  (\bibinfo{year}{2002}) \bibinfo{pages}{6538--6545}.
  \DOIprefix\doi{10.1063/1.1507116}.
\bibitem[{Han et~al.(2007)Han, Hu, Chen, Utkin, Brown, and
  Curl}]{han2007:Jetcooled}
\bibinfo{author}{J.~Han}, \bibinfo{author}{S.~Hu}, \bibinfo{author}{H.~Chen},
  \bibinfo{author}{Y.~Utkin}, \bibinfo{author}{J.~M. Brown},
  \bibinfo{author}{R.~F. Curl},
\newblock \bibinfo{title}{Jet-cooled infrared spectrum of methoxy in the {{CH}}
  stretching region},
\newblock \bibinfo{journal}{Phys. Chem. Chem. Phys.} \bibinfo{volume}{9}
  (\bibinfo{year}{2007}) \bibinfo{pages}{3725--3734}.
  \DOIprefix\doi{10.1039/b700502d}.
\bibitem[{Endo et~al.(1984)Endo, Saito, and
  Hirota}]{endo1984:Microwavespectrum}
\bibinfo{author}{Y.~Endo}, \bibinfo{author}{S.~Saito},
  \bibinfo{author}{E.~Hirota},
\newblock \bibinfo{title}{The microwave-spectrum of the methoxy radical
  {{CH}}{\textsubscript{3}}{{O}}},
\newblock \bibinfo{journal}{J. Chem. Phys.} \bibinfo{volume}{81}
  (\bibinfo{year}{1984}) \bibinfo{pages}{122--135}.
  \DOIprefix\doi{10.1063/1.447375}.
\bibitem[{Momose et~al.(1988)Momose, Endo, Hirota, and
  Shida}]{momose1988:Submillimeterwave}
\bibinfo{author}{T.~Momose}, \bibinfo{author}{Y.~Endo},
  \bibinfo{author}{E.~Hirota}, \bibinfo{author}{T.~Shida},
\newblock \bibinfo{title}{The submillimeter-wave spectrum of the
  ({{CH}}{\textsubscript{3}}{{O}})-{{C-13}} radical},
\newblock \bibinfo{journal}{J. Chem. Phys.} \bibinfo{volume}{88}
  (\bibinfo{year}{1988}) \bibinfo{pages}{5338--5343}.
  \DOIprefix\doi{10.1063/1.454593}.
\bibitem[{Laas and Weaver(2017)}]{laas2017:Millimeter}
\bibinfo{author}{J.~C. Laas}, \bibinfo{author}{S.~L.~W. Weaver},
\newblock \bibinfo{title}{The millimeter/submillimeter spectrum of the methoxy
  radical at low temperatures},
\newblock \bibinfo{journal}{Astrophys. J.} \bibinfo{volume}{835}
  (\bibinfo{year}{2017}) \bibinfo{pages}{46}.
  \DOIprefix\doi{10.3847/1538-4357/835/1/46}.
\bibitem[{Tsang and Hampson(1986)}]{tsang1986:Chemical}
\bibinfo{author}{W.~Tsang}, \bibinfo{author}{R.~F. Hampson},
\newblock \bibinfo{title}{Chemical {{Kinetic Data Base}} for {{Combustion
  Chemistry}}. {{Part I}}. {{Methane}} and {{Related Compounds}}},
\newblock \bibinfo{journal}{J. Phys. Chem. Ref. Data} \bibinfo{volume}{15}
  (\bibinfo{year}{1986}) \bibinfo{pages}{1087--1279}.
  \DOIprefix\doi{10.1063/1.555759}.
\bibitem[{Zellner et~al.(1986)Zellner, Fritz, and Lorenz}]{zellner1986:Methoxy}
\bibinfo{author}{R.~Zellner}, \bibinfo{author}{B.~Fritz},
  \bibinfo{author}{K.~Lorenz},
\newblock \bibinfo{title}{Methoxy formation in the reaction of
  {{CH}}{\textsubscript{3}}{{O}}{\textsubscript{2}} radicals with {{NO}}},
\newblock \bibinfo{journal}{J Atmos Chem} \bibinfo{volume}{4}
  (\bibinfo{year}{1986}) \bibinfo{pages}{241--251}.
  \DOIprefix\doi{10.1007/BF00052003}.
\bibitem[{Seinfeld and Pandis(2016)}]{seinfeld2016:Atmospheric}
\bibinfo{author}{J.~H. Seinfeld}, \bibinfo{author}{S.~N. Pandis},
  \bibinfo{title}{Atmospheric chemistry and physics: from air pollution to
  climate change}, \bibinfo{edition}{third edition} ed.,
  \bibinfo{publisher}{Wiley}, \bibinfo{address}{Hoboken, New Jersey},
  \bibinfo{year}{2016}.
\bibitem[{Anti{\~n}olo et~al.(2016)Anti{\~n}olo, Ag{\'u}ndez, Jim{\'e}nez,
  Ballesteros, Canosa, Dib, Albaladejo, and
  Cernicharo}]{antinolo2016:Reactivity}
\bibinfo{author}{M.~Anti{\~n}olo}, \bibinfo{author}{M.~Ag{\'u}ndez},
  \bibinfo{author}{E.~Jim{\'e}nez}, \bibinfo{author}{B.~Ballesteros},
  \bibinfo{author}{A.~Canosa}, \bibinfo{author}{G.~E. Dib},
  \bibinfo{author}{J.~Albaladejo}, \bibinfo{author}{J.~Cernicharo},
\newblock \bibinfo{title}{Reactivity of {{OH}} and
  {{CH}}{\textsubscript{3}}{{OH}} between 22 and 64 {{K}}: modeling the gas
  phase production of {{CH}}{\textsubscript{3}}{{O}} in {{Barnard}} 1b},
\newblock \bibinfo{journal}{Astrophys. J.} \bibinfo{volume}{823}
  (\bibinfo{year}{2016}) \bibinfo{pages}{25}.
  \DOIprefix\doi{10.3847/0004-637X/823/1/25}.
\bibitem[{Glasson(1975)}]{glasson1975:Methoxyl}
\bibinfo{author}{W.~A. Glasson},
\newblock \bibinfo{title}{Methoxyl radical reactions in atmospheric chemistry},
\newblock \bibinfo{journal}{Environ. Sci. Technol.} \bibinfo{volume}{9}
  (\bibinfo{year}{1975}) \bibinfo{pages}{1048--1053}.
  \DOIprefix\doi{10.1021/es60110a006}.
\bibitem[{Dames and Golden(2013)}]{dames2013:Master}
\bibinfo{author}{E.~E. Dames}, \bibinfo{author}{D.~M. Golden},
\newblock \bibinfo{title}{Master equation modeling of the unimolecular
  decompositions of hydroxymethyl ({{CH}}{\textsubscript{2}}{{OH}}) and methoxy
  ({{CH}}{\textsubscript{3}}{{O}}) radicals to formaldehyde
  ({{CH}}{\textsubscript{2}}{{O}}) + {{H}}},
\newblock \bibinfo{journal}{J. Phys. Chem. A} \bibinfo{volume}{117}
  (\bibinfo{year}{2013}) \bibinfo{pages}{7686--7696}.
  \DOIprefix\doi{10.1021/jp404836m}.
\bibitem[{Cernicharo et~al.(2012)Cernicharo, Marcelino, Roueff, Gerin,
  {Jimenez-Escobar}, and Munoz~Caro}]{cernicharo2012:CH3O}
\bibinfo{author}{J.~Cernicharo}, \bibinfo{author}{N.~Marcelino},
  \bibinfo{author}{E.~Roueff}, \bibinfo{author}{M.~Gerin},
  \bibinfo{author}{A.~{Jimenez-Escobar}}, \bibinfo{author}{G.~M. Munoz~Caro},
\newblock \bibinfo{title}{Discovery of the methoxy radical,
  {{CH}}{\textsubscript{3}}{{O}}, toward {{B1}}: {{Dust}} grain and gas-phase
  chemistry in cold dark clouds},
\newblock \bibinfo{journal}{Astrophys. J. Lett.} \bibinfo{volume}{759}
  (\bibinfo{year}{2012}) \bibinfo{pages}{L43}.
  \DOIprefix\doi{10.1088/2041-8205/759/2/L43}.
\bibitem[{Bacmann and Faure(2016)}]{bacmann2016:Origin}
\bibinfo{author}{A.~Bacmann}, \bibinfo{author}{A.~Faure},
\newblock \bibinfo{title}{The origin of gas-phase {{HCO}} and
  {{CH}}{\textsubscript{3}}{{O}} radicals in prestellar cores},
\newblock \bibinfo{journal}{Astron. Astrophys.} \bibinfo{volume}{587}
  (\bibinfo{year}{2016}) \bibinfo{pages}{A130}.
  \DOIprefix\doi{10.1051/0004-6361/201526198}.
\bibitem[{Ag{\'u}ndez et~al.(2019)Ag{\'u}ndez, Marcelino, Cernicharo, Roueff,
  and Tafalla}]{agundez2019:Sensitive}
\bibinfo{author}{M.~Ag{\'u}ndez}, \bibinfo{author}{N.~Marcelino},
  \bibinfo{author}{J.~Cernicharo}, \bibinfo{author}{E.~Roueff},
  \bibinfo{author}{M.~Tafalla},
\newblock \bibinfo{title}{A sensitive {\emph{{$\lambda$}}} 3 mm line survey of
  {{L483}}: {{A}} broad view of the chemical composition of a core around a
  {{Class}} 0 object},
\newblock \bibinfo{journal}{Astron. Astrophys.} \bibinfo{volume}{625}
  (\bibinfo{year}{2019}) \bibinfo{pages}{A147}.
  \DOIprefix\doi{10.1051/0004-6361/201935164}.
\bibitem[{{Guti{\'e}rrez-Quintanilla} et~al.(2021){Guti{\'e}rrez-Quintanilla},
  Layssac, Butscher, Henkel, Tsegaw, Grote, Sander, Borget, Chiavassa, and
  Duvernay}]{gutierrez-quintanilla2021:ICOM}
\bibinfo{author}{A.~{Guti{\'e}rrez-Quintanilla}}, \bibinfo{author}{Y.~Layssac},
  \bibinfo{author}{T.~Butscher}, \bibinfo{author}{S.~Henkel},
  \bibinfo{author}{Y.~A. Tsegaw}, \bibinfo{author}{D.~Grote},
  \bibinfo{author}{W.~Sander}, \bibinfo{author}{F.~Borget},
  \bibinfo{author}{T.~Chiavassa}, \bibinfo{author}{F.~Duvernay},
\newblock \bibinfo{title}{{{iCOM}} formation from radical chemistry: a
  mechanistic study from cryogenic matrix coupled with {{IR}} and {{EPR}}
  spectroscopies},
\newblock \bibinfo{journal}{Mon. Not. R. Astron. Soc.} \bibinfo{volume}{506}
  (\bibinfo{year}{2021}) \bibinfo{pages}{3734--3750}.
  \DOIprefix\doi{10.1093/mnras/stab1850}.
\bibitem[{Brown et~al.(1988)Brown, Charnley, and Millar}]{brown1988:Model}
\bibinfo{author}{P.~D. Brown}, \bibinfo{author}{S.~B. Charnley},
  \bibinfo{author}{T.~J. Millar},
\newblock \bibinfo{title}{A model of the chemistry in hot molecular cores},
\newblock \bibinfo{journal}{Mon. Not. R. Astron. Soc.} \bibinfo{volume}{231}
  (\bibinfo{year}{1988}) \bibinfo{pages}{409--417}.
  \DOIprefix\doi{10.1093/mnras/231.2.409}.
\bibitem[{Tielens and Whittet(1997)}]{tielens1997:Ices}
\bibinfo{author}{A.~Tielens}, \bibinfo{author}{D.~Whittet},
\newblock \bibinfo{title}{Ices in star forming regions},
\newblock \bibinfo{journal}{Symp. - Int. Astron. Union} \bibinfo{volume}{178}
  (\bibinfo{year}{1997}) \bibinfo{pages}{45--60}.
  \DOIprefix\doi{10.1017/S0074180900009232}.
\bibitem[{Watanabe and Kouchi(2002)}]{watanabe2002:Efficient}
\bibinfo{author}{N.~Watanabe}, \bibinfo{author}{A.~Kouchi},
\newblock \bibinfo{title}{Efficient {{Formation}} of {{Formaldehyde}} and
  {{Methanol}} by the {{Addition}} of {{Hydrogen Atoms}} to {{CO}} in
  {{H}}$_2${{O-CO Ice}} at 10~{{K}}},
\newblock \bibinfo{journal}{Astrophys. J.} \bibinfo{volume}{571}
  (\bibinfo{year}{2002}) \bibinfo{pages}{L173--L176}.
  \DOIprefix\doi{10.1086/341412}.
\bibitem[{Bermudez et~al.(2017)Bermudez, Bailleux, and
  Cernicharo}]{bermudez2017:Laboratory}
\bibinfo{author}{C.~Bermudez}, \bibinfo{author}{S.~Bailleux},
  \bibinfo{author}{J.~Cernicharo},
\newblock \bibinfo{title}{Laboratory detection of the rotational-tunnelling
  spectrum of the hydroxymethyl radical, {{CH}}{\textsubscript{2}} {{OH}}},
\newblock \bibinfo{journal}{A\&A} \bibinfo{volume}{598} (\bibinfo{year}{2017})
  \bibinfo{pages}{A9}. \DOIprefix\doi{10.1051/0004-6361/201629508}.
\bibitem[{Gerakines et~al.(1996)Gerakines, Schutte, and
  Ehrenfreund}]{gerakines1996}
\bibinfo{author}{P.~Gerakines}, \bibinfo{author}{W.~Schutte},
  \bibinfo{author}{P.~Ehrenfreund},
\newblock \bibinfo{title}{Ultraviolet processing of interstellar ice analogs:
  I. pure ices},
\newblock \bibinfo{journal}{Astronomy and Astrophysics} \bibinfo{volume}{312}
  (\bibinfo{year}{1996}) \bibinfo{pages}{289--305}.
\bibitem[{Watanabe et~al.(2007)Watanabe, Mouri, Nagaoka, Chigai, Kouchi, and
  Pirronello}]{watanabe2007:Laboratory}
\bibinfo{author}{N.~Watanabe}, \bibinfo{author}{O.~Mouri},
  \bibinfo{author}{A.~Nagaoka}, \bibinfo{author}{T.~Chigai},
  \bibinfo{author}{A.~Kouchi}, \bibinfo{author}{V.~Pirronello},
\newblock \bibinfo{title}{Laboratory {{Simulation}} of {{Competition}} between
  {{Hydrogenation}} and {{Photolysis}} in the {{Chemical Evolution}} of
  {{H}}{\textsubscript{2}} {{O}}-{{CO Ice Mixtures}}},
\newblock \bibinfo{journal}{ApJ} \bibinfo{volume}{668} (\bibinfo{year}{2007})
  \bibinfo{pages}{1001--1011}. \DOIprefix\doi{10.1086/521421}.
\bibitem[{Chuang et~al.(2016)Chuang, Fedoseev, Ioppolo, Van~Dishoeck, and
  Linnartz}]{chuang2016:Hatom}
\bibinfo{author}{K.-J. Chuang}, \bibinfo{author}{G.~Fedoseev},
  \bibinfo{author}{S.~Ioppolo}, \bibinfo{author}{E.~Van~Dishoeck},
  \bibinfo{author}{H.~Linnartz},
\newblock \bibinfo{title}{H-atom addition and abstraction reactions in mixed
  {{CO}}, {{H}}{\textsubscript{2}}{{CO}} and {{CH}}{\textsubscript{3}}{{OH}}
  ices -- an extended view on complex organic molecule formation},
\newblock \bibinfo{journal}{Mon. Not. R. Astron. Soc.} \bibinfo{volume}{455}
  (\bibinfo{year}{2016}) \bibinfo{pages}{1702--1712}.
  \DOIprefix\doi{10.1093/mnras/stv2288}.
\bibitem[{Tachikawa(1993)}]{tachikawa1993:Reaction}
\bibinfo{author}{H.~Tachikawa},
\newblock \bibinfo{title}{Reaction mechanism of the radical isomerization from
  {{CH}}{\textsubscript{3}}{{O}} to {{CH}}{\textsubscript{2}}{{OH}} in frozen
  methanol. {{An}} ab initio {{MO}} and {{RRKM}} study},
\newblock \bibinfo{journal}{Chem. Phys. Lett.} \bibinfo{volume}{212}
  (\bibinfo{year}{1993}) \bibinfo{pages}{27--31}.
  \DOIprefix\doi{10.1016/0009-2614(93)87102-9}.
\bibitem[{Cui and Morakuma(1996)}]{cui1996:Initio}
\bibinfo{author}{Q.~Cui}, \bibinfo{author}{K.~Morakuma},
\newblock \bibinfo{title}{Ab initio {{MO}} studies on the photodissociation of
  the methoxy family {{CX}}{\textsubscript{3}}{{Y}} ({{X}} = {{H}}, {{F}};
  {{Y}} = {{O}}, {{S}}) from the
  {{{\~A}}}{\textsuperscript{2}}{{A}}{\textsubscript{1}} state},
\newblock \bibinfo{journal}{Chemical Physics Letters} \bibinfo{volume}{263}
  (\bibinfo{year}{1996}) \bibinfo{pages}{54--62}.
  \DOIprefix\doi{10.1016/S0009-2614(96)01213-4}.
\bibitem[{Wang and Bowie(2010)}]{wang2010:Radical}
\bibinfo{author}{T.~Wang}, \bibinfo{author}{J.~H. Bowie},
\newblock \bibinfo{title}{Radical routes to interstellar glycolaldehyde.
  {{The}} possibility of stereoselectivity in gas-phase polymerization
  reactions involving {{CH}}{\textsubscript{2}}{{O}} and
  ˙{{CH}}{\textsubscript{2}}{{OH}}},
\newblock \bibinfo{journal}{Org. Biomol. Chem.} \bibinfo{volume}{8}
  (\bibinfo{year}{2010}) \bibinfo{pages}{4757}.
  \DOIprefix\doi{10.1039/c0ob00125b}.
\bibitem[{Burgers and Ruttink(2005)}]{burgers2005:Acidcatalyzed}
\bibinfo{author}{P.~C. Burgers}, \bibinfo{author}{P.~J. Ruttink},
\newblock \bibinfo{title}{The acid-catalyzed rearrangement
  {{CH}}{\textsubscript{3}}{{O}}{$\rightarrow$}{{CH}}{\textsubscript{2}}{{OH}}
  and its involvement in the dissociation of the methanol dimer radical
  cation},
\newblock \bibinfo{journal}{Int. J. Mass Spectrom.} \bibinfo{volume}{242}
  (\bibinfo{year}{2005}) \bibinfo{pages}{49--56}.
  \DOIprefix\doi{10.1016/j.ijms.2004.11.006}.
\bibitem[{Wang and Bowie(2012)}]{wang2012:Hydrogen}
\bibinfo{author}{T.~Wang}, \bibinfo{author}{J.~H. Bowie},
\newblock \bibinfo{title}{Hydrogen tunnelling influences the isomerisation of
  some small radicals of interstellar importance. {{A}} theoretical
  investigation},
\newblock \bibinfo{journal}{Org. Biomol. Chem.} \bibinfo{volume}{10}
  (\bibinfo{year}{2012}) \bibinfo{pages}{3219}.
  \DOIprefix\doi{10.1039/c2ob07102a}.
\bibitem[{Chitarra et~al.(2020)Chitarra, {Martin-Drumel}, Gans, Loison,
  Spezzano, Lattanzi, M{\"u}ller, and Pirali}]{chitarra2020:Reinvestigation}
\bibinfo{author}{O.~Chitarra}, \bibinfo{author}{M.-A. {Martin-Drumel}},
  \bibinfo{author}{B.~Gans}, \bibinfo{author}{J.-C. Loison},
  \bibinfo{author}{S.~Spezzano}, \bibinfo{author}{V.~Lattanzi},
  \bibinfo{author}{H.~S.~P. M{\"u}ller}, \bibinfo{author}{O.~Pirali},
\newblock \bibinfo{title}{Reinvestigation of the rotation-tunneling spectrum of
  the {{CH}}{\textsubscript{2}}{{OH}} radical: {{Accurate}} frequency
  determination of transitions of astrophysical interest up to 330 {{GHz}}},
\newblock \bibinfo{journal}{Astron. Astrophys.} \bibinfo{volume}{644}
  (\bibinfo{year}{2020}) \bibinfo{pages}{A123}.
  \DOIprefix\doi{10.1051/0004-6361/202039071}.
\bibitem[{Coudert et~al.(2022)Coudert, Chitarra, Spaniol, Loison,
  {Martin-Drumel}, and Pirali}]{coudert2022:CH2OH}
\bibinfo{author}{L.~H. Coudert}, \bibinfo{author}{O.~Chitarra},
  \bibinfo{author}{J.-T. Spaniol}, \bibinfo{author}{J.-C. Loison},
  \bibinfo{author}{M.-A. {Martin-Drumel}}, \bibinfo{author}{O.~Pirali},
\newblock \bibinfo{title}{Tunneling motion and splitting in the
  {CH{\textsubscript{2}}OH} radical: {(Sub-)millimeter} wave spectrum
  analysis},
\newblock \bibinfo{journal}{J. Chem. Phys.} \bibinfo{volume}{156}
  (\bibinfo{year}{2022}) \bibinfo{pages}{244301}.
  \DOIprefix\doi{10.1063/5.0095242}.
\bibitem[{{Cuadrado} et~al.(2015){Cuadrado}, {Goicoechea}, {Pilleri},
  {Cernicharo}, {Fuente}, and {Joblin}}]{Orion-Bar_mols_w_2-4C_2015}
\bibinfo{author}{S.~{Cuadrado}}, \bibinfo{author}{J.~R. {Goicoechea}},
  \bibinfo{author}{P.~{Pilleri}}, \bibinfo{author}{J.~{Cernicharo}},
  \bibinfo{author}{A.~{Fuente}}, \bibinfo{author}{C.~{Joblin}},
\newblock \bibinfo{title}{{The chemistry and spatial distribution of small
  hydrocarbons in UV-irradiated molecular clouds: the Orion Bar PDR}},
\newblock \bibinfo{journal}{Astron. Astrophys.} \bibinfo{volume}{575}
  (\bibinfo{year}{2015}) \bibinfo{pages}{A82}.
  \DOIprefix\doi{10.1051/0004-6361/201424568}.
  \href{http://arxiv.org/abs/1412.0417}{{\tt arXiv:1412.0417}}.
\bibitem[{{Cuadrado} et~al.(2017){Cuadrado}, {Goicoechea}, {Cernicharo},
  {Fuente}, {Pety}, and {Tercero}}]{Orion-Bar_COMs_2017}
\bibinfo{author}{S.~{Cuadrado}}, \bibinfo{author}{J.~R. {Goicoechea}},
  \bibinfo{author}{J.~{Cernicharo}}, \bibinfo{author}{A.~{Fuente}},
  \bibinfo{author}{J.~{Pety}}, \bibinfo{author}{B.~{Tercero}},
\newblock \bibinfo{title}{{Complex organic molecules in strongly UV-irradiated
  gas}},
\newblock \bibinfo{journal}{Astron. Astrophys.} \bibinfo{volume}{603}
  (\bibinfo{year}{2017}) \bibinfo{pages}{A124}.
  \DOIprefix\doi{10.1051/0004-6361/201730459}.
  \href{http://arxiv.org/abs/1705.06612}{{\tt arXiv:1705.06612}}.
\bibitem[{Chitarra et~al.(2022)Chitarra, Pirali, Spaniol, Hearne, Loison,
  Stanton, and {Martin-Drumel}}]{chitarra2022:CH2CN}
\bibinfo{author}{O.~Chitarra}, \bibinfo{author}{O.~Pirali},
  \bibinfo{author}{J.-T. Spaniol}, \bibinfo{author}{T.~S. Hearne},
  \bibinfo{author}{J.-C. Loison}, \bibinfo{author}{J.~F. Stanton},
  \bibinfo{author}{M.-A. {Martin-Drumel}},
\newblock \bibinfo{title}{{Pure rotational spectroscopy of the
  CH{\textsubscript{2}}CN radical extended to the sub-millimeter wave spectral
  region}},
\newblock \bibinfo{journal}{J. Phys. Chem. A} \bibinfo{volume}{126}
  (\bibinfo{year}{2022}) \bibinfo{pages}{7502--7513}.
  \DOIprefix\doi{10.1021/acs.jpca.2c04399}.
\bibitem[{{Pickett} et~al.(1998){Pickett}, {Poynter}, {Cohen}, {Delitsky},
  {Pearson}, and {M{\"u}ller}}]{JPL-catalog_1998}
\bibinfo{author}{H.~M. {Pickett}}, \bibinfo{author}{R.~L. {Poynter}},
  \bibinfo{author}{E.~A. {Cohen}}, \bibinfo{author}{M.~L. {Delitsky}},
  \bibinfo{author}{J.~C. {Pearson}}, \bibinfo{author}{H.~S.~P. {M{\"u}ller}},
\newblock \bibinfo{title}{{Submillimeter, millimeter and microwave spectral
  line catalog.}},
\newblock \bibinfo{journal}{J. Quant. Spectrosc. Radiat. Transfer}
  \bibinfo{volume}{60} (\bibinfo{year}{1998}) \bibinfo{pages}{883--890}.
  \DOIprefix\doi{10.1016/S0022-4073(98)00091-0}.
\bibitem[{{Pickett}(1991)}]{spfit_1991}
\bibinfo{author}{H.~M. {Pickett}},
\newblock \bibinfo{title}{{The fitting and prediction of vibration-rotation
  spectra with spin interactions}},
\newblock \bibinfo{journal}{J. Mol. Spectrosc.} \bibinfo{volume}{148}
  (\bibinfo{year}{1991}) \bibinfo{pages}{371--377}.
  \DOIprefix\doi{10.1016/0022-2852(91)90393-O}.
\bibitem[{{Martin-Drumel} et~al.(2012){Martin-Drumel}, {Eliet}, {Pirali},
  {Guinet}, {Hindle}, {Mouret}, and {Cuisset}}]{OH_SH_rot_2012}
\bibinfo{author}{M.~A. {Martin-Drumel}}, \bibinfo{author}{S.~{Eliet}},
  \bibinfo{author}{O.~{Pirali}}, \bibinfo{author}{M.~{Guinet}},
  \bibinfo{author}{F.~{Hindle}}, \bibinfo{author}{G.~{Mouret}},
  \bibinfo{author}{A.~{Cuisset}},
\newblock \bibinfo{title}{{New investigation on THz spectra of OH and SH
  radicals (X $^{2}{\Pi}_{i}$)}},
\newblock \bibinfo{journal}{Chem. Phys. Lett.} \bibinfo{volume}{550}
  (\bibinfo{year}{2012}) \bibinfo{pages}{8--14}.
  \DOIprefix\doi{10.1016/j.cplett.2012.08.027}.
\bibitem[{{Drouin}(2013)}]{drouin2013:OH}
\bibinfo{author}{B.~J. {Drouin}},
\newblock \bibinfo{title}{{Isotopic spectra of the hydroxyl radical}},
\newblock \bibinfo{journal}{J. Phys. Chem. A} \bibinfo{volume}{117}
  (\bibinfo{year}{2013}) \bibinfo{pages}{10076--10091}.
  \DOIprefix\doi{10.1021/jp400923z}.
\bibitem[{{M{\"u}ller} et~al.(2015){M{\"u}ller}, {Kobayashi}, {Takahashi},
  {Tomaru}, and {Matsushima}}]{NO_rot_2015}
\bibinfo{author}{H.~S.~P. {M{\"u}ller}}, \bibinfo{author}{K.~{Kobayashi}},
  \bibinfo{author}{K.~{Takahashi}}, \bibinfo{author}{K.~{Tomaru}},
  \bibinfo{author}{F.~{Matsushima}},
\newblock \bibinfo{title}{{Terahertz spectroscopy of N$^{18}$O and isotopic
  invariant fit of several nitric oxide isotopologs}},
\newblock \bibinfo{journal}{J. Mol. Spectrosc.} \bibinfo{volume}{310}
  (\bibinfo{year}{2015}) \bibinfo{pages}{92--98}.
  \DOIprefix\doi{10.1016/j.jms.2014.12.002}.
  \href{http://arxiv.org/abs/1412.4974}{{\tt arXiv:1412.4974}}.
\bibitem[{{Wong} et~al.(2017){Wong}, {Yurchenko}, {Bernath}, {M{\"u}ller},
  {McConkey}, and {Tennyson}}]{NO_fitting_2017}
\bibinfo{author}{A.~{Wong}}, \bibinfo{author}{S.~N. {Yurchenko}},
  \bibinfo{author}{P.~{Bernath}}, \bibinfo{author}{H.~S.~P. {M{\"u}ller}},
  \bibinfo{author}{S.~{McConkey}}, \bibinfo{author}{J.~{Tennyson}},
\newblock \bibinfo{title}{{ExoMol line list - XXI. Nitric Oxide (NO)}},
\newblock \bibinfo{journal}{Mon. Not. R. Astron. Soc.} \bibinfo{volume}{470}
  (\bibinfo{year}{2017}) \bibinfo{pages}{882--897}.
  \DOIprefix\doi{10.1093/mnras/stx1211}.
  \href{http://arxiv.org/abs/1705.05955}{{\tt arXiv:1705.05955}}.
\bibitem[{{Miller} and {Cohen}(2001)}]{IO_rot_2001}
\bibinfo{author}{C.~E. {Miller}}, \bibinfo{author}{E.~A. {Cohen}},
\newblock \bibinfo{title}{{Rotational spectroscopy of IO X $^{2}{\Pi}_{i}$}},
\newblock \bibinfo{journal}{J. Chem. Phys.} \bibinfo{volume}{115}
  (\bibinfo{year}{2001}) \bibinfo{pages}{6459--6470}.
  \DOIprefix\doi{10.1063/1.1398308}.
\bibitem[{{Anttila} et~al.(1993){Anttila}, {Horneman}, {Koivusaari}, and
  {Paso}}]{MeCN_DeltaK=3_1993}
\bibinfo{author}{R.~{Anttila}}, \bibinfo{author}{V.~M. {Horneman}},
  \bibinfo{author}{M.~{Koivusaari}}, \bibinfo{author}{R.~{Paso}},
\newblock \bibinfo{title}{{Ground state constants $A_{0}$, $D^{K}_{0}$ and
  $H^{K}_{0}$ of CH$_{3}$CN}},
\newblock \bibinfo{journal}{J. Mol. Spectrosc.} \bibinfo{volume}{157}
  (\bibinfo{year}{1993}) \bibinfo{pages}{198--207}.
  \DOIprefix\doi{10.1006/jmsp.1993.1016}.
\bibitem[{{Graner}(1976)}]{graner1976:CH3F}
\bibinfo{author}{G.~{Graner}},
\newblock \bibinfo{title}{{Determination of $A_0$ for CH$_3$F from ground-state
  combination differences}},
\newblock \bibinfo{journal}{Mol. Phys} \bibinfo{volume}{31}
  (\bibinfo{year}{1976}) \bibinfo{pages}{1833--1843}.
  \DOIprefix\doi{10.1080/00268977600101441}.
\bibitem[{{M{\"u}ller} et~al.(2015){M{\"u}ller}, {Brown}, {Drouin}, {Pearson},
  {Kleiner}, {Sams}, {Sung}, {Ordu}, and {Lewen}}]{MeCN_v8le2_2015}
\bibinfo{author}{H.~S.~P. {M{\"u}ller}}, \bibinfo{author}{L.~R. {Brown}},
  \bibinfo{author}{B.~J. {Drouin}}, \bibinfo{author}{J.~C. {Pearson}},
  \bibinfo{author}{I.~{Kleiner}}, \bibinfo{author}{R.~L. {Sams}},
  \bibinfo{author}{K.~{Sung}}, \bibinfo{author}{M.~H. {Ordu}},
  \bibinfo{author}{F.~{Lewen}},
\newblock \bibinfo{title}{{Rotational spectroscopy as a tool to investigate
  interactions between vibrational polyads in symmetric top molecules:
  Low-lying states $\varv_{8} \leq 2$ of methyl cyanide, CH$_{3}$CN}},
\newblock \bibinfo{journal}{J. Mol. Spectrosc.} \bibinfo{volume}{312}
  (\bibinfo{year}{2015}) \bibinfo{pages}{22--37}.
  \DOIprefix\doi{10.1016/j.jms.2015.02.009}.
  \href{http://arxiv.org/abs/1502.06867}{{\tt arXiv:1502.06867}}.
\bibitem[{{M{\"u}ller} et~al.(2021){M{\"u}ller}, {Belloche}, {Lewen}, {Drouin},
  {Sung}, {Garrod}, and {Menten}}]{MeCN_up2v4eq1_etc_2021}
\bibinfo{author}{H.~S.~P. {M{\"u}ller}}, \bibinfo{author}{A.~{Belloche}},
  \bibinfo{author}{F.~{Lewen}}, \bibinfo{author}{B.~J. {Drouin}},
  \bibinfo{author}{K.~{Sung}}, \bibinfo{author}{R.~T. {Garrod}},
  \bibinfo{author}{K.~M. {Menten}},
\newblock \bibinfo{title}{{Toward a global model of the interactions in
  low-lying states of methyl cyanide: Rotational and rovibrational spectroscopy
  of the {\ensuremath{\upsilon}}$_{4}$ = 1 state and tentative interstellar
  detection of the {\ensuremath{\upsilon}}$_{4}$ =
  {\ensuremath{\upsilon}}$_{8}$ = 1 state in Sgr B2(N)}},
\newblock \bibinfo{journal}{J. Mol. Spectrosc.} \bibinfo{volume}{378}
  (\bibinfo{year}{2021}) \bibinfo{pages}{111449}.
  \DOIprefix\doi{10.1016/j.jms.2021.111449}.
  \href{http://arxiv.org/abs/2103.07389}{{\tt arXiv:2103.07389}}.
\bibitem[{{M{\"u}ller}(2013)}]{PH3_analysis_2013}
\bibinfo{author}{H.~S.~P. {M{\"u}ller}},
\newblock \bibinfo{title}{{Spectroscopic parameters of phosphine, PH$_{3}$, in
  its ground vibrational state}},
\newblock \bibinfo{journal}{J. Quant. Spectrosc. Radiat. Transfer}
  \bibinfo{volume}{130} (\bibinfo{year}{2013}) \bibinfo{pages}{335--340}.
  \DOIprefix\doi{10.1016/j.jqsrt.2013.05.002}.
  \href{http://arxiv.org/abs/1305.1138}{{\tt arXiv:1305.1138}},
  \bibinfo{note}{and references therein.}
\bibitem[{{M{\"u}ller} et~al.(2001){M{\"u}ller}, {Thorwirth}, {Roth}, and
  {Winnewisser}}]{CDMS_2001}
\bibinfo{author}{H.~S.~P. {M{\"u}ller}}, \bibinfo{author}{S.~{Thorwirth}},
  \bibinfo{author}{D.~A. {Roth}}, \bibinfo{author}{G.~{Winnewisser}},
\newblock \bibinfo{title}{{The Cologne Database for Molecular Spectroscopy,
  CDMS}},
\newblock \bibinfo{journal}{Astron. Astrophys.} \bibinfo{volume}{370}
  (\bibinfo{year}{2001}) \bibinfo{pages}{L49--L52}.
  \DOIprefix\doi{10.1051/0004-6361:20010367}.
\bibitem[{{M{\"u}ller} et~al.(2005){M{\"u}ller}, {Schl{\"o}der}, {Stutzki}, and
  {Winnewisser}}]{CDMS_2005}
\bibinfo{author}{H.~S.~P. {M{\"u}ller}}, \bibinfo{author}{F.~{Schl{\"o}der}},
  \bibinfo{author}{J.~{Stutzki}}, \bibinfo{author}{G.~{Winnewisser}},
\newblock \bibinfo{title}{{The Cologne Database for Molecular Spectroscopy,
  CDMS: a useful tool for astronomers and spectroscopists}},
\newblock \bibinfo{journal}{J. Mol. Struct.} \bibinfo{volume}{742}
  (\bibinfo{year}{2005}) \bibinfo{pages}{215--227}.
  \DOIprefix\doi{10.1016/j.molstruc.2005.01.027}.
\bibitem[{{Endres} et~al.(2016){Endres}, {Schlemmer}, {Schilke}, {Stutzki}, and
  {M{\"u}ller}}]{CDMS_2016}
\bibinfo{author}{C.~P. {Endres}}, \bibinfo{author}{S.~{Schlemmer}},
  \bibinfo{author}{P.~{Schilke}}, \bibinfo{author}{J.~{Stutzki}},
  \bibinfo{author}{H.~S.~P. {M{\"u}ller}},
\newblock \bibinfo{title}{{The Cologne Database for Molecular Spectroscopy,
  CDMS, in the Virtual Atomic and Molecular Data Centre, VAMDC}},
\newblock \bibinfo{journal}{J. Mol. Spectrosc.} \bibinfo{volume}{327}
  (\bibinfo{year}{2016}) \bibinfo{pages}{95--104}.
  \DOIprefix\doi{10.1016/j.jms.2016.03.005}.
  \href{http://arxiv.org/abs/1603.03264}{{\tt arXiv:1603.03264}}.
\bibitem[{{Jackels}(1982)}]{CH3O_ai_dip_etc_1982}
\bibinfo{author}{C.~F. {Jackels}},
\newblock \bibinfo{title}{{A theoretical potential energy surface study of
  several states of the methoxy radical}},
\newblock \bibinfo{journal}{J. Chem. Phys.} \bibinfo{volume}{76}
  (\bibinfo{year}{1982}) \bibinfo{pages}{505--515}.
  \DOIprefix\doi{10.1063/1.442752}.
\bibitem[{{Ulenikov} et~al.(1999){Ulenikov}, {Onopenko}, {Tyabaeva},
  {Schroderus}, and {Alanko}}]{CH3D_v0_1999}
\bibinfo{author}{O.~N. {Ulenikov}}, \bibinfo{author}{G.~A. {Onopenko}},
  \bibinfo{author}{N.~E. {Tyabaeva}}, \bibinfo{author}{J.~{Schroderus}},
  \bibinfo{author}{S.~{Alanko}},
\newblock \bibinfo{title}{{On the rotational analysis of the ground vibrational
  state of CH$_{3}$D molecule}},
\newblock \bibinfo{journal}{J. Mol. Spectrosc.} \bibinfo{volume}{193}
  (\bibinfo{year}{1999}) \bibinfo{pages}{249--259}.
  \DOIprefix\doi{10.1006/jmsp.1998.7729}.
\bibitem[{Chahbazian et~al.(2024)Chahbazian, {Martin-Drumel}, and
  Pirali}]{chahbazian2024:CH2CHO}
\bibinfo{author}{R.~Chahbazian}, \bibinfo{author}{M.-A. {Martin-Drumel}},
  \bibinfo{author}{O.~Pirali},
\newblock \bibinfo{title}{{High-resolution spectroscopic investigation of the
  CH{\textsubscript{2}}CHO radical in the sub-millimeter region}},
\newblock \bibinfo{journal}{J. Phys. Chem. A} \bibinfo{volume}{128}
  (\bibinfo{year}{2024}) \bibinfo{pages}{370--377}.
  \DOIprefix\doi{10.1021/acs.jpca.3c06326}.
\bibitem[{Martin-Drumel et~al.(2024)Martin-Drumel, Coutens, Loison, Jørgensen,
  and Pirali}]{martin-drumel2024:H2NCO}
\bibinfo{author}{M.-A. Martin-Drumel}, \bibinfo{author}{A.~Coutens},
  \bibinfo{author}{J.-C. Loison}, \bibinfo{author}{J.~K. Jørgensen},
  \bibinfo{author}{O.~Pirali},
\newblock \bibinfo{title}{Unveiling gas phase \ce{H2NCO} radical -- laboratory
  rotational spectroscopy and interstellar searches},
\newblock \bibinfo{journal}{Astronom. Astrophys.} \bibinfo{volume}{687}
  (\bibinfo{year}{2024}) \bibinfo{pages}{A233}.
  \DOIprefix\doi{10.1051/0004-6361/202449711}.
\bibitem[{Chahbazian et~al.(2024)Chahbazian, Juppet, and
  Pirali}]{chahbazian2024:Faraday}
\bibinfo{author}{R.~Chahbazian}, \bibinfo{author}{L.~Juppet},
  \bibinfo{author}{O.~Pirali},
\newblock \bibinfo{title}{Unveiling the spectroscopy of complex organic
  radicals by exploiting faraday rotation at (sub-)millimeter wavelengths.
  {Illustration} with the acetonyl radical},
\newblock \bibinfo{journal}{J. Phys. Chem. Lett.} \bibinfo{volume}{15}
  (\bibinfo{year}{2024}) \bibinfo{pages}{9803--9810}.
  \DOIprefix\doi{10.1021/acs.jpclett.4c01936}.

\end{thebibliography}

\end{document}